\pdfoutput=1
\documentclass{article}
\usepackage{arxiv}
\usepackage{multirow}
\usepackage{amsbsy}
\usepackage[utf8]{inputenc}
\usepackage[T1]{fontenc}
\usepackage{bm}
\usepackage{enumitem}

\usepackage{hyperref}       
\usepackage{url}            

\usepackage{amsfonts}       
\usepackage{nicefrac}       
\usepackage{microtype}      
\usepackage[square,numbers]{natbib}
\usepackage{doi}
\usepackage{booktabs}

\usepackage{array}
\newcolumntype{P}[1]{>{\centering\arraybackslash}p{#1}}

\usepackage{tikz}
\usetikzlibrary{matrix}
\usetikzlibrary{chains,positioning}
\usetikzlibrary{decorations.pathreplacing,angles,quotes}
\usetikzlibrary{shapes,arrows}
\usetikzlibrary{calc}
\usetikzlibrary{fit}

\usepackage{pifont}
\usepackage{longtable}
\usepackage{rotating}

\tikzset{empty node/.style={draw=none,fill=none}}
\makeatletter
\def\tikz@lib@matrix@empty@cell{\iftikz@lib@matrix@empty\node[name=\tikzmatrixname-\the\pgfmatrixcurrentrow-\the\pgfmatrixcurrentcolumn,empty node]{};\fi}
\makeatother

\newcommand{\npr}{\mathit{npr}}
\newcommand{\maxnpr}{\mathit{maxnpr}}
\newcommand{\anpr}{\mathit{anpr}}
\newcommand{\nprv}{\mathit{nprv}}

\newcommand{\nnz}{\mathit{nnz}}

\definecolor{armygreen}{rgb}{0.29, 0.33, 0.13}

\definecolor{britishracinggreen}{rgb}{0.0, 0.26, 0.15}

\definecolor{ao(english)}{rgb}{0.0, 0.5, 0.0}

\definecolor{maroon(x11)}{rgb}{0.69, 0.19, 0.38}

\newcommand{\stfo}{\textcolor{blue}{Stfo}}

\newcommand{\hierarchical}{\textcolor{blue}{Hierarchical}}

\newcommand{\stfocm}{\textcolor{blue}{Coalesced-Memory}}

\newcommand{\stfolb}{\textcolor{blue}{Load-Balancing}}
\newcommand{\stfoma}{\textcolor{blue}{Memory-Alignment}}
\newcommand{\stfotd}{\textcolor{blue}{Thread-Divergence}}
\newcommand{\stfotm}{\textcolor{blue}{Thread-Mapping}}
\newcommand{\bandw}{\textcolor{blue}{Bandwidth-Specific}}
\newcommand{\devdev}{\textcolor{blue}{Device-Device}}
\newcommand{\mem}{\textcolor{blue}{Memory-Specific}}

\newcommand{\singlef}{\textcolor{blue}{Single-Format}}

\newcommand{\cofo}{\textcolor{ao(english)}{Cofo}}

\newcommand{\dacache}{\textcolor{ao(english)}{Cache-Specific}}
\newcommand{\dacompiler}{\textcolor{ao(english)}{Compiler-Specific}}

\newcommand{\cofoat}{\textcolor{ao(english)}{Auto-Tuning}}
\newcommand{\cofolb}{\textcolor{ao(english)}{Load-Balancing}}
\newcommand{\cofobuf}{\textcolor{ao(english)}{Buffering}}
\newcommand{\cofocm}{\textcolor{ao(english)}{Coalesced}}
\newcommand{\cofoma}{\textcolor{ao(english)}{Memory-Alignment}}
\newcommand{\cofotd}{\textcolor{ao(english)}{Thread-Divergence}}
\newcommand{\cofotm}{\textcolor{ao(english)}{Thread-Mapping}}
\newcommand{\cofor}{\textcolor{ao(english)}{Reduction}}
\newcommand{\cofodr}{\textcolor{ao(english)}{Data-Reuse}}
\newcommand{\cofodl}{\textcolor{ao(english)}{Data-Locality}}

\newcommand{\auto}{\textcolor{maroon(x11)}{Auto-Selective}}

\newcommand{\mls}{\textcolor{maroon(x11)}{ML-Supervised}}
\newcommand{\dls}{\textcolor{maroon(x11)}{DL-Supervised}}

\newcommand{\matr}[1]{\bm{#1}}     %

\newcommand{\etal}{\textit{et al}.}
\newcommand{\ie}{\textit{i}.\textit{e}.}
\newcommand{\eg}{\textit{e}.\textit{g}.}

\usepackage{subcaption}

\title{Performance Enhancement Strategies for Sparse Matrix-Vector Multiplication (SpMV) and Iterative Linear Solvers}

\author{ \href{https://orcid.org/0000-0002-4767-4147}{\includegraphics[scale=0.06]{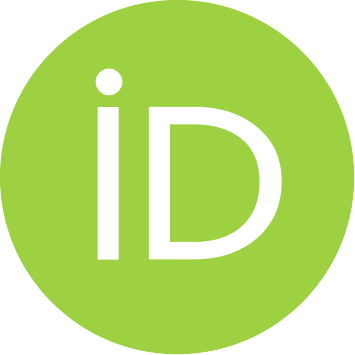}\hspace{1mm}Thaha~Mohammed}\\
  Department of Computer Science\\
  Aalto University\\
  Espoo, Finland 02150 \\
  \texttt{thaha.mohammed@aalto.fi} \\
\And
  \href{https://orcid.org/0000-0002-4997-5322}{\includegraphics[scale=0.06]{orcid.pdf}\hspace{1mm}Rashid Mehmood} \\
  High Performance Computing Center\\
  King Abdulaziz University\\
  Jeddah, Saudi Arabia 23185 \\
  \texttt{RMehmood@kau.edu.sa} \\
}

\renewcommand{\headeright}{}
\renewcommand{\undertitle}{}
\renewcommand{\shorttitle}{Performance Enhancement Strategies for SpMV}

\hypersetup{
pdftitle={Performance Enhancement Strategies for Sparse Matrix-Vector Multiplication (SpMV) and Iterative Linear Solvers},
pdfsubject={q-bio.NC, q-bio.QM},
pdfauthor={Thaha Mohammed, Rashid Mehmood},
pdfkeywords={Sparse Matrix, Sparse storage, Survey, Linear iterative Solvers, GPU}
}

\setlist[description]{font=\itshape}

\date{}
\begin{document}

\maketitle

\begin{abstract}
Iterative solutions of sparse linear systems and sparse eigenvalue problems have a fundamental role in vital fields of scientific research and engineering. The crucial computing kernel for such iterative solutions is the multiplication of a sparse matrix by a dense vector. Efficient implementation of sparse matrix-vector multiplication (SpMV) and linear solvers are therefore essential and has been subjected to extensive research across a variety of computing architectures and accelerators such as central processing units (CPUs), graphical processing units (GPUs), many integrated cores (MICs), and field programmable gate arrays (FPGAs). Unleashing the full potential of an architecture/accelerator requires determining the factors that affect an efficient implementation of SpMV. This article presents the first of its kind, in-depth survey covering over two hundred state-of-the-art optimization schemes for solving sparse iterative linear systems with a focus on computing SpMV. A new taxonomy for iterative solutions and SpMV techniques common to all architectures is proposed. This article includes reviews of SpMV techniques for all architectures to consolidate a single taxonomy to encourage cross-architectural and heterogeneous-architecture developments. However, the primary focus is on GPUs. The major contributions as well as the primary, secondary, and tertiary contributions of the SpMV techniques are first highlighted utilizing the taxonomy and then qualitatively compared. A summary of the current state of the research for each architecture is discussed separately. Finally, several open problems and key challenges for future research directions are outlined.\end{abstract}

\keywords{Sparse Matrix, Sparse Storage, Sparse Matrix-Vector Multiplication (SpMV), Linear iterative Solvers, Central Processing Units (CPUs), Graphical Processing Units (GPUs), Many Integrated Cores (MICs), Field Programmable Gate Arrays (FPGAs), Survey}

\section{Introduction}
\label{sec:Introduction}

Linear algebra is vital to many fields of science and engineering. Especially sparse linear algebra, which is included by the Berkeley scientists in their set of motifs (the seven dwarfs~\cite{Asanovic:EECS-2006-183}). Among the sparse numerical techniques, iterative solution of linear systems can be considered as of prime importance due to its application in various important areas such as solving finite differences of partial differential equations (PDEs)~\cite{Golovashkin:2013:Sfef}, high accuracy surface modelling~\cite{Yan:2015:Suth:S}, finding steady-state probability vector of Markov chains~\cite{Mehmood:APIM:MASCOTS}, solving the time-fractional Schr\"{o}dinger equation~\cite{Garrappa:2015:SttS:JCP}, web ranking~\cite{Langville:2005:ASoE:S}, inventory control and manufacturing systems~\cite{Cinneide:1994:Smom}, queuing systems~\cite{Kim:2015:Assq:PEVA,Ching:2013:MaRS}, fault modelling, weather forecasting, stochastic automata networks~\cite{Alex:2020:ASAN}, communication systems~\cite{Heffes:1986:AMMC:JSAC}, reliability analysis, wireless networks~\cite{Bylina:2012:AMMo},
sensor networks~\cite{Park:2009:AgMc:MOBHOC}, computational biology~\cite{Bustamam:2012:FPMC:TCBB}, computational physics, and natural
language modelling~\cite{lafferty2001conditional}.

Sparse matrix operations forms the basis for sparse linear systems. Any matrix in which the fraction of non-zero elements to the total elements is considerably low is a sparse matrix. A matrix $\matr{A} \in R^{m\times n}$, is a sparse matrix when  $\nnz \ll m \times n$, where $\nnz$ is the number of non-zero elements in $\matr{A}$. Sparse matrices that arise from real life problems are large but consist of a
small number of nonzero elements. Hence, they require specialized storage schemes and algorithms to efficiently store, access, and compute the sparse matrices.

Sparse linear systems are generally of the form $\matr{A}\matr{x}=\matr{b}$, where $\matr{A}$ is a sparse matrix and $\matr{b}$, $\matr{x}$ are dense vectors. The solution of the system that needs to be obtained is $\matr{x}$. Such system of linear equations can be solved by direct methods or iterative methods~\cite{Varga:2000:MIA,Owe:2004:Ism:CRMM}. The direct methods are based on factorizations as seen in libraries such as LAPACK~\cite{Anderson:1999:LAPACK} and SuiteSparse~\cite{Kolodziej:2019:TSMC}. Gaussian elimination, LU, Cholesky and QR~\cite{Owe:2004:Ism:CRMM} factorization are the major direct methods. As the size of the system increase the direct methods become expensive due to fill-in phenomenon; wherein new non-zero entries, in addition to existing ones, are generated in matrix $\matr{A}$. For \eg{}, when a row of a sparse matrix is subtracted from another row, some of the zero entries in the latter row may become nonzero. Such modifications to the matrix mean that the data structure employed to store the sparse matrix must be updated during the execution of the algorithm thus resulting in growing memory and slower performance. 

Whereas, iterative methods do not alter the matrix $A$, they only use the matrix in the context of SpMV~\cite{Mehmood:2004:SDAo}; hence there is a need to develop an efficient sparse storage and associated implementation of optimized sparse matrix-vector multiplication. Iterative solvers generate approximate solutions for linear systems by successive approximations at each iteration~\cite{Barret:1995:TftS,Mehmood:APIM:MASCOTS,Mehmood:2004:SDAo}. They start out with an approximate solution and iteratively modify the approximation until convergence. However, they do not guarantee a convergence, but they are faster than the direct methods. The iterative methods can be further classified into stationary (iterative function does not change across iteration) iterative methods~\cite{Varga:2000:MIA,Bai:2003:HaSS:S} like Jacobi and Gauss-Seidl, and non-stationary iterative methods~\cite{saad2013iterative,Owe:2004:Ism:CRMM} including popular Krylov subspace techniques such as Conjugate Gradient (CG) and Generalized Minimal Residual (GMRES). The volume of literature on general iterative methods are huge~\cite{Barret:1995:TftS,Owe:2004:Ism:CRMM,H:1993:Scai:CHOICE,saad2013iterative}. A survey on the existing iterative solutions for linear systems can be found in~\cite{Saad:2000:Isol}.

Optimization of SpMV performance poses a significant challenge on any architecture. SpMV is heavily memory bound; i.e., the ratio of global memory access to floating point operations is high due to indirect memory access. Moreover, the bandwidth utilization is dependent on the sparsity structure of input matrix and the computing platform(s) utilized. Hence, the challenge in developing efficient and compact data structures suited to each architecture and improving the bandwidth efficiency. Optimization and the performance improvement of single threaded applications have given way to parallel computing to attain a higher throughput. The vanguards among the current massively parallel chip-based architectures are the GPUs (Graphics Processing Units), MICs (Many Integrated Cores), and FPGAs (Field Programmable Gate Arrays). 

The advent  of GPUs as a revolutionary technology providing massive parallelism and higher throughput has resulted in the accelerated performance of scientific applications. A large number of scientific applications has been parallelized to run on GPU for improved performance. Moreover, the ubiquity of the technology enables researchers easy access to GPUs. The key elements for running serial algorithms on GPU are the design, analysis, and implementation of parallel algorithms that scale to hundreds of coupled cores. The difficulty is in adapting the serial code to parallel architecture. GPU is a throughput device. Utilizing the full potential of GPU requires exposing a large amount of fine-grained parallelism, computations should be structured such that there is regularity in the execution path and the memory access. 

MIC architecture (also known as Intel Xeon Phi Knights Landing) is a highly parallel engine and efficient processor architecture that achieve high performance through utilization of
large of number of core like vector registers and high bandwidth hierarchical memory. MIC supports the use of AVX-512 instruction set and extensions. Cores in MIC supports quad-directional simultaneous multithreading (SMT) and is accompanied by vector processing units with which branching can be avoided among the threads; this can be exploited by applications with irregular and unaligned access. With a large number of cores, one critical performance obstacle of SpMV on KNC is memory boundedness of SpMV and load imbalance.

FPGAs provide a hardware solution using fine-grained parallelism which is a cost effective means to achieve a high performance. Besides the high floating-point performance, the current FPGAs  come with large amounts of on-chip memory. These features enable FPGA designs to provide high on-chip and off-chip memory bandwidth to I/O-bound applications such as SpMV~\cite{Underwood:CtGC:FCCM}. Moreover, FPGAs consume lesser power as compared to GPUs and MICs.

SpMV implementations for CPUs, GPUs, and MICs focus on absolving issues related to the dissonance between the memory access pattern of the sparse matrices (due to irregular structure) and the SIMD architecture; hence the development of the sparse storage scheme as an “interface” or “bridge” to improve the performance.

This article proposes taxonomies for the classification of linear solver enhancement techniques with a focus on SpMV. Based on the proposed taxonomy, sparse storage formats and associated SpMV techniques with a focus on GPUs are discussed. The major contributions as well as the primary, secondary, and tertiary contributions of the SpMV techniques for GPUs are first highlighted utilizing the taxonomy and then qualitatively compared. Furthermore, we also qualitatively discuss the same for CPU, MIC, and FPGA based implementations. A summary of current state of the research for each architecture is discussed separately. Finally, several open problems and key challenges for future research directions are outlined. To the best of our knowledge, this survey is the first of its kind in this area of the investigation developing a taxonomy across various architectures. Filippone \etal{}~\cite{Filippone:2017:SMMo} review SpMV storage formats for GPU and classify sparse storage based on the well known base formats. Any storage format not based on these base formats are not classified. Moreover,~\cite{Filippone:2017:SMMo} is largely focused on quantitative experiments. In contrast, our taxonomy is based on both the storage and compute aspects and focuses on the primary, secondary, and tertiary functionalities of the data structure and provide extensive qualitative comparisons. A taxonomy at such level of details allows to see various architectural, data structure and algorithmic features that various schemes have used and how these relate to designing the best schemes in terms of memory access patterns, throughput, bandwidth efficiency, and load balancing.

The rest of the paper is as follows. Section~\ref{sec:taxonomy} proposes a taxonomy of SpMV performance enhancement techniques for sparse iterative solvers and general SpMV\@. Section~\ref{sec:basic} discuss the fundamental storage schemes common to GPUs, CPUs, MICs, and FPGAs. Later, taxonomy becomes clearer in Section~\ref{sec:gpu} and Section~\ref{sec:cpu} where more than one hundred and fifty state of the art schemes for GPUs and CPUs, are discussed and analyzed respectively, using the taxonomy. Section~\ref{sec:fpga} and Section~\ref{sec:mic} discuss and analyze the state of the art schemes for FPGAs and MICs respectively. Furthermore, Section~\ref{sec:hetro} utilizes the taxonomy to discuss the SpMV techniques and various libraries for SpMV operation on heterogeneous architectures. Finally, Section~\ref{sec:conclusion} 
summarizes our observations on current state and trends in improving SpMV computations. Furthermore, it outlines current research issues and future research directions for SpMV computations.

\begin{figure}[t]
  \centerline{\includegraphics[scale=0.56]{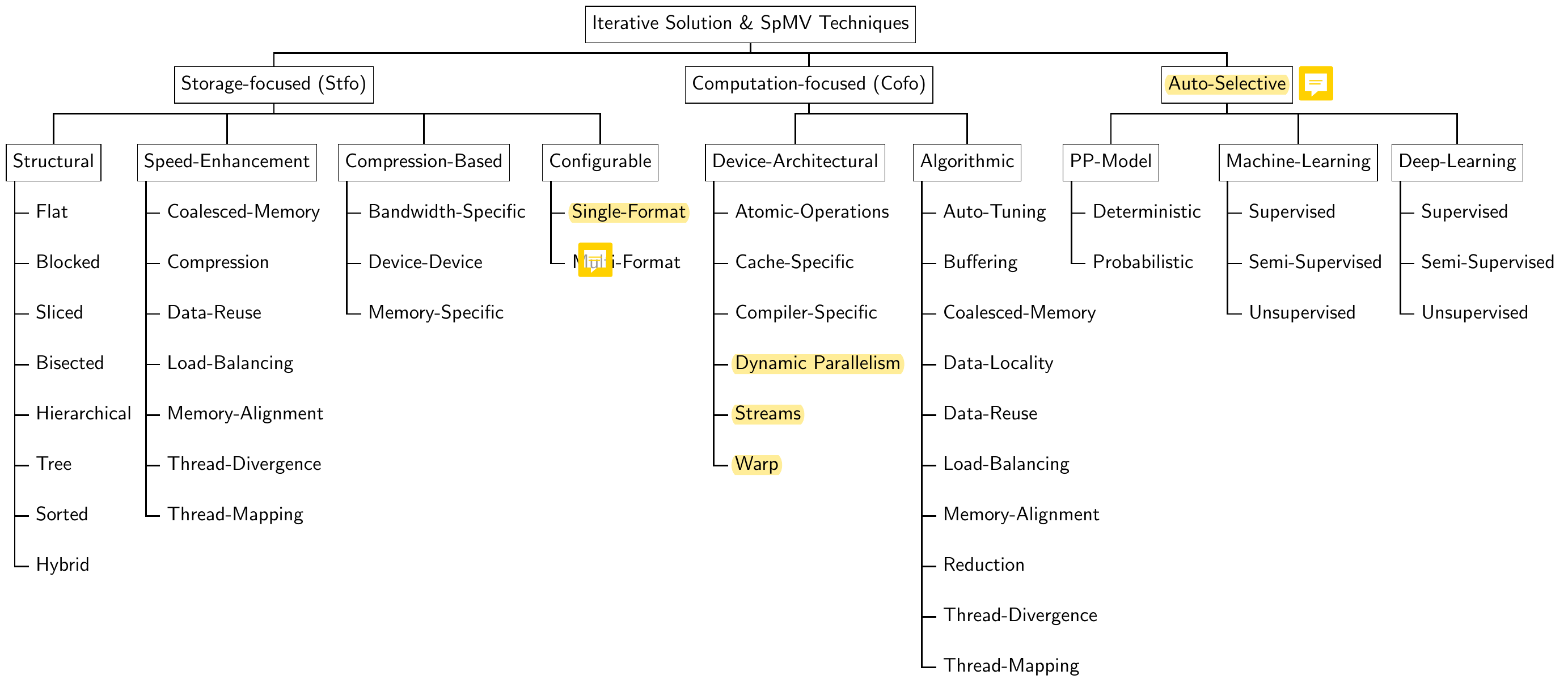}}
  \caption{Taxonomy of SpMV techniques for iterative solution of large sparse linear equation systems.}\label{fig:taxo}
\end{figure}

\section{SpMV and Iterative Solution Techniques: The Taxonomy}
\label{sec:taxonomy}

This section proposes a taxonomy (see Figure~\ref{fig:taxo}) of performance enhancement techniques for SpMV computations on the major high performance computing architectures (CPU, GPU, FPGA, MIC, and heterogeneous architectures). The review and taxonomy also includes the works on iterative solution of large sparse linear equation systems that have a focus on SpMV performance. 
The classifications in the taxonomy are described briefly in the rest of this section. Further elaborations will come as the individual schemes and their taxonomies are discussed in the subsequent Sections~\ref{sec:gpu} to ~\ref{sec:hetro}.
An SpMV scheme may address a single or multiple performance characteristics (\eg, reducing storage requirements as well as enhancing speed) and hence the schemes are mainly discussed based on their primary foci. The secondary and tertiary classifications are also given wherever applicable; see \eg, Table~\ref{tab:gpu:summ}.
We classify the SpMV techniques into three broad categories: (1) Storage-focused (Stfo), (2) Computation-focused (Cofo), and (3) Auto-Selective (Sefo) techniques (See Level 1 taxonomy in Figure~\ref{fig:taxo}). 

\subsection{Definitions: Storage-focused (Stfo) Techniques}
\label{sec:taxonomy:stfo}

Stfo techniques are those which ``primarily focus on developing data structures to store a sparse matrix with the aim to reduce memory requirements or data transfer, or improve speed of SpMV computations''. These are classified into four categories (see Figure~\ref{fig:taxo}): Structural, Speed-Enhancement, Compression-Based, and Configurable techniques (See Level 2 taxonomy on the left side of Figure~\ref{fig:taxo}).

\textbf{Structural} techniques store a matrix in a particular structure such as a Block. Figure~\ref{fig:taxo} lists the various sparse scheme structures that have been reported in the literature. These include Flat, Blocked, Sliced (or Segmented), Bisected, Hierarchical, Tree, Sorted, and Hybrid formats.
\textbf{Flat} storage structures or formats store the row and/or column indices of the nonzero elements explicitly (\eg, CSR~\cite{saad1994sparskit}, ELLPACK~\cite{GrimesKincaidYoung1979}). 
\textbf{Blocked} formats (\eg, BCSR~\cite{Choi:2010:Maos}, BCCOO~\cite{YanLiZhangEtAl2014,Zhang:2016:ACSF}, BCMSR~\cite{UCAM-CL-TR-650}) divide a sparse matrix into blocks and store them separately along with the block metadata. 
\textbf{Sliced} structures (\eg, SCOO~\cite{Dang:2012:TSCF:PROCS}, SELLPACK~\cite{Alex:2010:ATSM}) are in fact Blocked structures except that in this case a block spans the whole row; \ie, a matrix is divided into blocks, called slices, where each block or slice contains multiple (complete) rows of the matrix. 
\textbf{Bisected} formats (\eg, BiELL~\cite{Zheng:2014:BAbE:JPDC}) are hybrid of Blocked and Sliced formats, where each slice is further divided into blocks. 
\textbf{Tree} formats (\eg, MQT~\cite{Simecek:2012:MQFf:SYNASC}) store the nonzero elements and their indices in a tree structures. The indices of the elements can be explicit or implicit. Quadtree~(\cite{Simecek:2009:SMCU:SYNASC,zhang2016efficient,Simecek:2012:MQFf:SYNASC}) and Binary Tree~(\cite{Balk:2004:BBST}) are subclasses of the Tree format. 
\textbf{Hierarchical} formats (\eg, BBCS~\cite{Stathis2004}, COOCSR~\cite{LangrSimecekTvrdik2013}) use two or more levels to store a sparse matrix. A particular level in the hierarchy may store the nonzeros, or both nonzeros and their indices. 
\textbf{Sorted} formats (\eg,~SELL-P~\cite{anzt2014implementing}) store matrix rows or columns by ascending or descending order of the number of nonzeroes (npr).
\textbf{Hybrid} formats (\eg,~\cite{Alex:2009:IBSM}) divide a matrix into parts and store each part using a different storage scheme.

\textbf{Speed-Enhancement} techniques are sparse storage scheme that enhance the speed of sparse computations (\eg, throughput measured in FLOPS). These can be classified based on the specific {\em performance characteristics} of the computations that they are able to enhance. The performance characteristics include \textbf{Coalesced-Memory} (combining multiple memory accesses into a single transaction)~\cite{Yang:2012:Aism:IPEMC}, 
\textbf{Compression} (compressing data to improve speed)~\cite{Tang:2015:AFoB:TPDS},  
\textbf{Data-Reuse} (reusing data and reducing memory accesses)~\cite{Maggioni:2013:AAAW:ICPP}, 
 \textbf{Load-balancing} (balancing load across multiple threads, processes, etc.)~\cite{Yan:2014:Mboo:S}, \textbf{Memory Alignment} (aligned memory access to the matrix and the vector elements)~\cite{Vuduc:2005:OAlo}, 
\textbf{Thread-Divergence} (avoiding threads diverging from a common execution trajectory)~\cite{Maggioni:2013:AATf:PROCS}, and \textbf{Thread-Mapping} (mapping threads to execution resources, kernels, etc.)~\cite{Yin:2013:POfS}. Note that these categories of works propose {\em storage schemes} that primarily, or otherwise, attempt to improve computing speed.

\textbf{Compression-based} techniques use various compression methods to reduce memory or bandwidth requirements, classified into \textbf{Bandwidth-Specific} (compression to reduce the internal memory bandwidth requirements of a processor)~\cite{Yan:2014:Mboo:S}, \textbf{Device-Device} (compression to reduce bandwidth usage and transfer time between devices, CPUs, GPUs, FPGAs, MICs, disks, etc.)~\cite{Wieczorek:2014:AEWo}, or \textbf{Memory-Specific} (compression to reduce memory requirements)~\cite{Zhang:2016:ACSF}.
The Bandwidth-Specific schemes are {\em primarily} designed to improve the internal memory bandwidth efficiency to address memory-bound nature of SpMV computations (bandwidth and irregular data access). 
The Device-Device schemes use compression to reduce bandwidth requirements between processors and devices such as between GPU and host CPU or CPU and disk.
The Memory-Specific schemes are able to solve larger matrices for a given hardware by compression and, also, could typically result in faster computations due to improved caching and memory bandwidth efficiency.

\textbf{Configurable} techniques dynamically construct a storage format at run-time based on certain sparse matrix features with the aim to provide improved performance. These could be \textbf{Single-Format} (atmost a single format is used to configure the storage scheme)~\cite{liu2015csr5} or \textbf{Multi-Format} (two or more formats are used to configure the storage scheme)~\cite{Zhang:2016:ACSF}. To elaborate further, these techniques may configure matrix block sizes~\cite{Choi:2010:Maos,Choi:2010:Maos} or slice sizes, etc.,  (AdELL~\cite{Maggioni:2013:AAAW:ICPP}) to improve performance. 
Other types of configurable techniques may appear in the future (using machine learning, for example).

\subsection{Definitions: Computation-focused (Cofo) Techniques}\label{sec:taxonomy:cofo}

\textbf{Computation-focused (Cofo)} techniques (see Figure~\ref{fig:taxo}) {\em ``improve SpMV performance by utilizing the general design or individual constructs provided by the hardware or instruction set architecture (ISA) ({\normalfont called} \textbf{Device-Architectural}), or by developing better algorithms ({\normalfont called }\textbf{Algorithmic})''}. 
\textbf{Device-Architectural} techniques may exploit an optimized construct for atomic operations offered by a device manufacturer to improve performance (\textbf{Atomic-Operations}~\cite{Grewe:2011:Agat}), or use other constructs to improve performance such as related to cache operations (\textbf{Cache-Specific}~\cite{Mukunoki:2013:OoSM}), or compiler options (\textbf{Compiler-Specific}~\cite{Grewe:2011:Agat}), or related to dynamic parallelism (\textbf{Dynamic Parallelism}~\cite{Muhammed:2019:SANM:APP}), or Streams in GPUs (\textbf{Streams-Specific}~\cite{Guo:2013:Aots}), or GPU
 warps (\textbf{Warp-Specific}~\cite{barbieri2015three}).  

\textbf{Algorithmic} techniques could be classified based on their focus on \textbf{Auto-Tuning} (tuning various parameters to improve performance such as by improved caching)~\cite{Whaley:1998:ATLA:SC}, \textbf{Buffering} (buffering to improve processor access latency to nonzeros)~\cite{Bol:2011:Ombu}, \textbf{Coalesced-Memory}~\cite{Wijs:2012:IGSM}, \textbf{Data-Locality} (reducing data movement)~\cite{Li:2016:Adld}, \textbf{Data-Reuse}~\cite{Alappat:2020:ARAC}, \textbf{Load-Balancing}~\cite{Flegar:2017:OLIf}, \textbf{Memory-Alignment}~\cite{Chen:2020:tAtl:INS}, \textbf{Reduction} (algorithmic improvements to the reduction operation in SpMV)~\cite{Mukunoki:2013:OoSM}, \textbf{Thread-Divergence}~\cite{Ashari:2014:FSMM:SC}, and \textbf{Thread-Mapping}~\cite{Maggioni:2013:AAAW:ICPP}. 
Some of the Cofo classes above
have already been defined earlier in Section~\ref{sec:taxonomy:stfo}.

\subsection{Definitions: Auto-Selective (Sefo) Techniques}
\label{sec:taxonomy:sefo}

Auto-Selective (Sefo) techniques (see Figure~\ref{fig:taxo}) {\em ``aim to select an optimal sparse storage format and/or SpMV computational kernel for a given sparse matrix from a set of formats and/or kernels to store and/or compute SpMV''}. The selection of storage format or kernel is based on methods including performance prediction models (PP-Model), machine learning, and deep learning. 
The method types and classes can be extended in the future with other AI methods. 

\textbf{PP-Model} techniques can be \textbf{Deterministic} (\ie~the model is deterministic)~\cite{Guo:2014:Acpm:CPE} or \textbf{Probabilistic} (uses a probabilistic model)~\cite{Li:2013:S} to predict performance of a set of sparse formats or kernels for the solution of a given matrix and decide accordingly. Sefo schemes based on PP models are hard to build and extend as it is necessary to build a specific model for each sparse format under consideration. The current PP models make multiple assumptions to simplify the underlying hardware complexities.

\textbf{Machine-Learning} techniques can be \textbf{Supervised} (\eg~\cite{Elafrou:2019:B}), \textbf{Semi-Supervised}, or \textbf{Unsupervised}. A similar classification of \textbf{Deep-Learning} techniques is possible (see Figure~\ref{fig:taxo}). These classes will evolve with new techniques expected to be proposed in the future.
Note that Sefo and Auto-Tuning (see Section~\ref{sec:taxonomy:cofo}) techniques by definition are different. Auto-Tuning techniques select SpMV parameters (sparsity or device architectural features, etc.) to optimise performance while Sefo techniques select a storage scheme or compute kernel. Future schemes are expected to have a mix of fine-grained (autotuning) and course-grained (Sefo) selection.

\begin{table}
\centering
\caption{Important Notations}
\label{tab:def}
\begin{tabular}{ll}
\toprule
Symbols & Definition\\ 
\midrule
$n$           &  Row size of a matrix\\ 
$m$           &  Column size of a matrix\\
$\nnz$        &  Total number of non zeroes in a matrix\\
$\npr$        &  Vector of non zeroes per row in a matrix\\
$\anpr$       &  Average non zeroes per row in a matrix\\
$\maxnpr$     &  The highest non-zeros per row among all rows\\
$\nprv$       &  The variance of non zeroes per row among all rows\\
\bottomrule
\end{tabular}
\end{table}

\section{Fundamental Storage Schemes (All Platforms)}
\label{sec:basic}

We review here the fundamental sparse storage schemes that have been in use since the early days of sparse computations on all platforms including CPUs, FPGAs, GPUs, and MICs. 

\subsection{Coordinate Format (COO)}\label{sec:basic:coo}
\begin{figure}[tbh]
\begin{subfigure}{0.4\columnwidth}
    \centering
    \includegraphics[scale=.76]{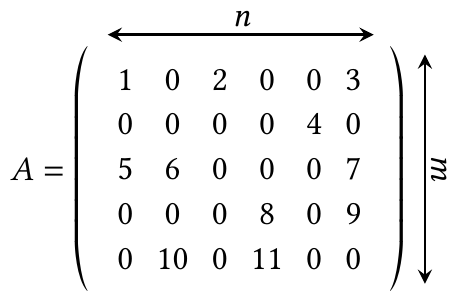}\caption{}\label{fig:matrix-A}\end{subfigure}
  \hfill \begin{subfigure}{0.55\columnwidth}
    \includegraphics[scale=.76]{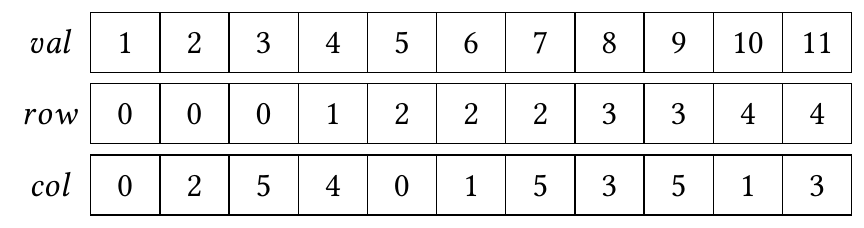}\caption{}\label{fig:coo}\end{subfigure}\\
  \begin{subfigure}{0.4\columnwidth}
    \includegraphics[scale=.76]{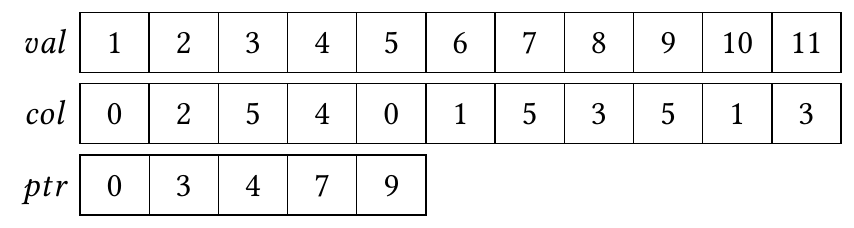}\caption{}\label{fig:csr}\end{subfigure}
\hfill \begin{subfigure}{0.5\columnwidth}
  \centering
    \includegraphics[scale=.76]{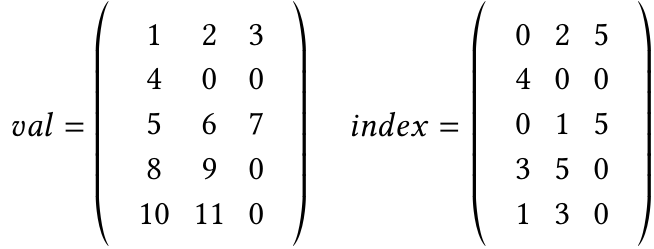}\caption{}\label{fig:ell}\end{subfigure}
    \caption{Representations of matrix \(A\) as:~(\subref{fig:matrix-A}) Dense, (\subref{fig:coo}) COO, (\subref{fig:csr}) CSR, and (\subref{fig:ell}) ELL\label{fig:sparse} storage schemes}
\end{figure}

The coordinate (COO) format~\cite{saad1994sparskit} (Figure~\ref{fig:coo}) is the most basic sparse storage scheme. It uses three arrays: $val$, $row$, and $col$, each of length corresponding to the number of non-zeroes ($\nnz$), to store the matrix nonzero elements and their respective row and column indices, respectively. 
Given an 8-byte floating point number representation and a 4-byte integer representation, the COO format requires $16\times \nnz$ bytes to store the whole sparse matrix. 

\subsection{Compressed Sparse Row (CSR)}\label{sec:basic:csr}

The compressed sparse row (CSR) format (Figure~\ref{fig:csr}), similar to COO, stores the nonzero values and their respective column indices in arrays $val$ and $col$ (of length $\nnz$), respectively.
A third array, $ptr$, is used to point to the beginning of the each rows. For an $m\times n$ matrix, $ptr$ of length $m+1$ stores the offset of the $i^{th}$ row in $ptr\left[i\right]$; $ptr\left[m \right]$ stores $\nnz$. 
The CSR format requires $12\times nnz+4\times(m+1)$ bytes for storage. 
Generally, one thread is assigned per row for SpMV operation and this technique is known as CSR scalar. To improve the performance one warp is assigned to a row and this technique is called CSR vector~\cite{BellGarland2008}.

\subsection{ELLPACK (ELL)}\label{sec:basic:ellpack}

ELLPACK~\cite{GrimesKincaidYoung1979} also known as ITPACK is a sparse
matrix storage format well-suited for traditional CPUs as well as vector architectures such as GPUs and MICs.
The sparse matrix $A$ is stored using the following data structures (Figure~\ref{fig:ell}): (a) A $2D$ floating point
array, $val$, of size $m\times \maxnpr$, 
to store the nonzero values. The rows that contain nonzeros less than $\maxnpr$ (\ie, $npr[i] < \maxnpr$) are zero-padded. (b) A $2D$ integer array, $col$, of size $m\times \maxnpr$ to store the respective column indices. Again the rows with $npr[i] < \maxnpr$, are zero-padded. The ELL format requires $12\times m\times \maxnpr$ bytes to store a sparse matrix. Evidently, the storage efficiency of ELLPACK is affected for matrices with high $\nprv$ value (\ie, $\maxnpr >> \anpr$). 

\subsection{Diagonal Storage (DIA)}\label{sec:basic:dia}

Diagonal format is only useful for storing sparse matrices with a diagonal structure.
It uses two arrays, a $val$ array for storing the values and an \textit{offset} array that stores the diagonal offsets of the values from the main diagonal. The main diagonal has offset zero. A positive offset indicates super-diagonals and a negative offset indicates sub-diagonals. The size of the \textit{offset} array would be equal to the number of nonzero diagonals of the matrix. 

\section{SpMV Techniques on {GPUs}}\label{sec:gpu}

This section discuss the SpMV techniques for GPUs in detail utilizing the discussed taxonomy.

\subsection{Storage-focused (Stfo) Techniques (GPU{s})}\label{sec:gpu:stfo}
\subsubsection{Structural Techniques}\label{sec:gpu:stfo:structural}
\begin{description}[leftmargin=0pt, itemsep=0.5em, topsep=0pt]
\item[Flat.]\label{sec:gpu:stfo:structural:flat}
The Coordinate (COO) format, the most basic sparse storage scheme, was originally introduced for CPUs~\cite{saad1994sparskit} and later implemented for GPUs in~\cite{BellGarland2008}. It uses three arrays: $val$, $row$, and $col$, each of length $\nnz$ (see Table~\ref{tab:def}), to store the matrix nonzero elements and their respective row and column indices, respectively. 
Compressed Sparse Row (CSR)~(CPUs~\cite{saad1994sparskit}, GPUs~\cite{BellGarland2008}), similar to COO, stores the nonzeros and their respective column indices in arrays $val$ and $col$ (of length $\nnz$), respectively. The difference lies in the third array, $ptr$, which is used to point to the beginning of the each row. For an $m\times n$ matrix, $ptr$ of length $m+1$ stores the offset of the $i$th row in $ptr\left[i\right]$; $ptr\left[m \right]$ stores $\nnz$. It is also known as Compressed Row Storage (CRS). Compressed Sparse Column (CSC) or Compressed Column Storage (CCS) stores a matrix by columns. 
Generally, in GPU implementations, one thread is assigned per row for SpMV operation and this technique is known as CSR Scalar. To improve the performance one warp is assigned to a row and this technique is called CSR Vector~\cite{BellGarland2008, eguly:2012:Esmm:INPAR}.
ELLPACK, 
also known as ELL or ITPACK, is well-suited for traditional CPUs~\cite{GrimesKincaidYoung1979} as well as vector architectures such as GPUs~\cite{BellGarland2008} and MICs. 
The sparse matrix $\matr{A}$ is stored using the following data structures. (a) A $2$D floating point array, $val$, of size $m\times \maxnpr$, to store the nonzero values. The rows that contain nonzeros less than $\maxnpr$ (\ie, $npr[i] < \maxnpr$) are zero-padded. (b) A $2D$ integer array, $col$, of size $m\times \maxnpr$ to store the respective column indices. Again the rows with $npr[i] < \maxnpr$, are zero-padded. The ELL format requires $12\times m\times \maxnpr$ bytes to store a sparse matrix. Apart from CSR, ELL and other basic flat schemes, the JAD sparse storage format (also called JDS)~\cite{Li:2012:Gpil:S,Cevahir:2009:FCGw} was among the earliest flat formats to be adopted from CPU~\cite{Saad:1989:KSMo} to GPUs. JAD can be considered as an optimization of the ELL format on GPUs. Initial preprocessing is required to store a sparse matrix in JAD. The sparse input array is sorted initially based on $\npr$
in the descending order. After sorting, the zeroes are removed from the array and the nonzeros are shifted left. The columns of the newly formed structure are called the Jagged diagonals. Four arrays are used to store a matrix in JAD; a floating point array that stores the nonzeros in the column-major order, and three integer arrays to store the corresponding column indices, the pointers to the jagged diagonals, and the row permutations. JAD provides better SpMV performance than ELL, however the GPU occupancy obtained in JAD is lesser than ELL. 
The Diagonal (DIA) format~(CPUs~\cite{saad1994sparskit}, GPUs~\cite{BellGarland2008}) is only useful for storing sparse matrices with a diagonal structure. It uses two arrays, a $val$ array for storing the values and an $\mathit{offset}$ array that stores the diagonal index offsets of the values from the main diagonal. The main diagonal has offset zero. A positive offset indicates super-diagonals and a negative offset indicates sub-diagonals. The size of the $\mathit{offset}$ array would be equal to the number of nonzero diagonals of the matrix. 

Oberhuber \etal~\cite{OberhuberSuzukiVacata2010} proposed the Row grouped CSR (RgCSR) storage scheme based on CSR and Sliced ELLPACK in order to improve coalesced memory access on GPUs. Initially, the matrix is divided into groups of a certain number of rows. For each group, the first element from each row is stored followed by the second element in each row until the last element in each row. If the number of elements in the rows of a group is not equal, they are padded with zeroes along with the column indices. Each of the group is assigned to a CUDA block and each row is accessed by a thread. They claim to reduce thread divergence and improve coalesced memory access. 
Dehnavi \etal~\cite{Dehnavi:2010:FSMV:TMAG} proposed the Prefetch Compressed CSR (PCSR) format, based on CSR, for the acceleration of finite element solvers, wherein they use new partitioning and computing scheme with zero-padding that enables pre-fetching of data. 
ELL-R was introduced in~\cite{Vazquez:2010:Anaf:CPE} to improve the performance of the ELL format on GPUs. In addition to the ELL data structures it has an integer array to store the length of the each row. This additional data structure improves the performance of SpMV by reducing thread divergence within a warp as the row length information enables the skipping of nonzeros. The total number of rows ($m$) has to be a divisor of the block size for coalesced global memory access and hence extra rows with zero values are padded to achieve this. If the new row size after padding is $m'$, then the memory required to store ELL-R format for double-precision matrices is $2mk \times 8 + m' \times 4$ bytes.
The ELLR-T format~\cite{Vazquez:2012:Atot:PARCO} is based on GPU kernel modification of ELL-R~\cite{Vazquez:2010:Anaf:CPE}. Unlike ELLR, where each row is handled by a thread during SpMV, ELLR-T, uses a $T$ number of threads to operate on the row where the $T$ can be 1, 2, 4, 8, 16, or 32. A preprocessing is performed on ELL-R, including permutation of the nonzeros and their column indices, and adding zero padding so that each row is a multiple of 16, also ensuring coalesced memory access. 
Each row in the value array is split into a $T$ parts, and $T$ threads compute the SpMV. Partial results are stored in the shared memory and the final results are obtained by reduction in shared memory.
Sun \etal~\cite{Sun:2011:OSfD:ICPP} propose a storage format called Compressed Row Segment with Diagonal-pattern (CRSD) to reduce the zero padding of DIA. The diagonals are divided into multiple groups based on the diagonal pattern to deal with the idle sections. Moreover, they split the matrix into a number of row segments. In each segment, the nonzeros on the diagonals of the same group are considered as a unit of storage and operation. The nonzeros are stored contiguously and the operations on them are performed together. 
Yang \etal~\cite{Yang:2012:Aism:IPEMC} propose Improved CSR (ICSR), an improvement to CSR, aimed at SpMV on CUDA-based GPUs. The authors solve the issue of serialization in the CSR storage caused due to $\nprv$ by padding the rows with zeroes which results in aligned access to the global memory. However, ICSR increases the space complexity and causes the threads to iterate over extra elements that degrades the performance.
Liu and Vinter~\cite{Liu:2015:Sssf:PARCO} proposed a new SpMV storage scheme and algorithm based on the CSR storage format using speculative segmented sums. A segmented sum operation is done on the GPU to generate possibly correct results and later the partial sums are rearranged on the CPU to obtain the correct result vector. The proposed algorithm mainly consists of two steps: the speculative segment execution stage on the GPU and the prediction checking stage on the CPU. Segmented sum techniques have also been covered earlier on CPUs~\cite{blelloch1993segmented} and GPUs~\cite{SenguptaHarrisZhangEtAl2007,Dotsenko:2008:Fsao}. 
Other Flat schemes include~\cite{Merrill:2016:Msmm}

\item[Blocked.]\label{sec:gpu:stfo:structural:blocked} 
Buatois \etal~\cite{Buatois:2009:Cnca} proposed an implementation of BCRS (BCRS/CRS is another name for BCSR/CSR) where the high-level block structure of the matrix is stored using CSR and the matrix blocks are stored as dense matrices to leverage the massive parallelism provided by GPUs. 
Note that this blocked implementation is different from fully sparse implementation of sparse formats (see e.g.~\cite{UCAM-CL-TR-650}) where both the matrix blocks as well as the location of these blocks are stored using sparse formats. The dense blocks store both nonzero and zeros of the block and hence may have higher memory requirements than fully sparse implementation.
Choi \etal~\cite{Choi:2010:Maos,Choi:2010:Maos} implemented a blocked version of CSR (BCSR) for GPUs and a blocked version of ELL, the blocked ELLPACK (BELLPACK). A matrix $m\times n$ in CSR is converted into BCSR by partitioning the matrix into $(m/r)\times(n/c)$ sub-blocks of size $r\times c$. For creating the BELLPACK format initially the matrix is converted into sub-blocks similar to the BCSR. Then the block rows are sorted depending upon the number of blocks per row, after which the matrix is subdivided into smaller matrices and each of these smaller matrices are stored in the ELLPACK format.
Yang \etal~\cite{Yang:2011:Fsmm} improved the performance of SpMV for Data Mining applications that follow power law characteristics. They transform the matrix and use tiling to improve the temporal locality. This is achieved by reordering each column by their lengths and tiling the input matrix partially.
Godwin \etal~\cite{Godwin:2012:Hsmm} proposed Column Diagonal Storage (CDS), which takes advantage of the diagonals in the matrices. They were motivated by the challenges in SpMV computations for structured grid problems that have block structures where techniques such as DIA or CSR are inefficient. Storing the matrix based on block diagonal structure improves coalesced memory access and lowers the memory requirements due to lower zero-padding compared to DIA.
Abu-Sufah and Abdel Karim~\cite{Abu-Sufah:2012:AEAf:HPCC} extended the TJDS/TJAD format using blocking to propose Blocked Transpose Jagged Diagonal Storage (BTJDS). The matrix is divided into blocks and each block is stored in the TJDS format. The authors assign a fixed number of warps to a single block. Each column in a matrix is assigned a thread, and they iterate until all columns are traversed. 
Xu et al.~\cite{Xu:2012:OSMV:SNPD} proposed a cache blocking method for optimizing SpMV on GPUs. The input matrix is partitioned into multiple sub-blocks and each sub-block is stored in CSR. When the column size of a sub-block is small then it can be reused in the GPU cache. One row of blocks is considered as a CUDA block, and threads are assigned per row for each row of sub-matrix blocks. Ashari \etal~\cite{Ashari:2014:Aetb} proposes the Blocked Row-Column (BRC) format that utilizes two dimensional blocking for the storage of sparse matrices. The rows are sorted and reordered and combined into various groups. The rows are partitioned into blocks with each block containing a constant number of nonzeros per block. The row sparse matrices are permuted like JDS and are combined with the ELL storage mechanism. Padding is done for each row inside the blocks to reduce the padding compared to ELL. Moreover, the sparse matrix is again blocked along the column dimension. This results in the scheme being adaptive to the structural characteristics of the matrix. 
See also the BIN-BCSR scheme~\cite{Weber:2012:EGDS:J} in Section~``\nameref{sec:gpu:stfo:structural:sorted}'' on Page~\pageref{sec:gpu:stfo:structural:sorted}.
Liu and Vinter~\cite{liu2015csr5} proposed CSR5. 
CSR5 uses a combination of row block methods~\cite{Ashari:2014:FSMM:SC} and segmented sum methods~\cite{blelloch1993segmented, SenguptaHarrisZhangEtAl2007,Dotsenko:2008:Fsao}, and is shown to provide fairly consistent performance for $23$ matrices with different sparsity patterns and $1$ dense matrices on GPUs, CPUs, and MICs.
There are a number of other schemes (\eg~\cite{Alex:2009:IBSM, Kreutzer:2012:SMMo:IPDPSW, Nagasaka:2016:AMBO:PROCS, YanLiZhangEtAl2014,Zhang:2016:ACSF}) that are based on blocked storage of matrices. These are mentioned in other sections because their primary focus or storage method relate to other categories.

\item[Sliced.]\label{sec:gpu:stfo:structural:sliced}
Monakov \etal~\cite{Alex:2010:ATSM} proposed Sliced ELLPACK (SELLPACK) to reduce zero-padding needed by ELL\@. The matrix is partitioned into a variable number of slices with each slice consisting of a variable number of adjacent rows. Each slice is stored separately using ELL\@. The slicing reduces the difference between the shortest and the longest row resulting in lower zero-padding. One CUDA thread block can be assigned to each slice for CUDA implementation. 
Dziekonski et al.~\cite{Dziekonski:2011:AMEA:PIER} proposed Sliced ELLR-T scheme, designed specifically to solve complex-valued sparse linear equations in computational electromagnetics. Sliced ELLR-T is based on Sliced ELL and ELLR-T. It uses fixed sized slicing of Sliced ELL and multiple threads ($T= 1, 2, 4, 8, 16, 32$) for a row as seen in ELLR-T that provides a higher performance due to coalesced memory access. Sliced ELLR-T requires smaller memory than ELL-R and ELLR-T but larger than CSR.
Feng et al.~\cite{Feng:2011:OoSM:ICPADS} proposed the Segmented Interleave Combination (SIC) format based on CSR. For a sparse matrix with rows $m$, $c$ contiguous rows are selected and combined to form a new row. Only one warp and a global memory access are required to perform a single multiplication and addition. The SIC format is sorted according to the row length and converted into segments and each segment can also be stored using SIC format and different GPU kernels can be launched for each segment. Each warp processes $c$ rows. SIC hence reduces the thread divergence and improves coalesced memory access.
Dang and Schmidt et al.~\cite{Dang:2012:TSCF:PROCS} proposed Sliced COO (SCOO) and shown that it performs better than the COO implementation of cuSparse~\cite{BellGarland2008} and the CUSP library~\cite{Cusp}. Initially, the matrix is sorted based on the row weights. The values are then stored in column major format ensuring that the elements in the same columns are contiguously located. The local nonzero elements are sorted inside each row and the slicing is done using consecutive rows. The SpMV kernel utilizes one CUDA block for each slice. An atomic add operation is used to add the results in shared memory. The authors also proposed a faster heuristic based partitioning algorithm for the matrix A into SCOO format. The memory required by SCOO is less than HYB and COO but more than CSR.
Maggioni et al.~\cite{Maggioni:2013:GSSo:IPDPSW} proposed improvements to SELLPACK~\cite{Alex:2010:ATSM} by selecting the slices equal to the warp size and performing local reordering within each slice to balance the load without affecting the cache locality. This technique reduces the overhead associated with SELLPACK~\cite{Alex:2010:ATSM} and provides comparatively better performance.
Koza et al.~\cite{Koza:2014:CMSF} proposed compressed multi-row sparse format (CMRS). The input matrix is split into slices of fixed and equal sizes. It is a sliced version of CSR and the conversion between CSR and CMRS is trivial. It dynamically assigns threads per row and each strip is executed by a SIMD unit. The number of rows executed by a warp is not constant in this format.
Barbieri et al.~\cite{barbieri2015three} proposed Hacked ELLPACK and Hacked DIA to address the limitations of ELLPACK and DIA respectively. Hacked ELLPACK solves the issue of the extra zero padding required for ELL. They divide the matrix into slices (called hacks) and store these hacks separately.  The hacks are designed to be a multiple of the warp size. Hacked DIA is similar to Hacked ELL, with the difference that the hacks are stored using DIA instead of ELLPACK. They also propose an optimization by improving the ELLPACK based SpMV that is similar to ELLPACK-R~\cite{Vazquez:2010:Anaf:CPE}. 
Other schemes in this category include~\cite{YanLiZhangEtAl2014,Zhang:2016:ACSF, Hou:2017:ASfP:IPDPSW, Kreutzer:2014:AUSM, Benatia:2019:Smpf}.

\item[Bisected.]\label{sec:gpu:stfo:structural:bisected}
Bisection ELLPACK (BiELL)~\cite{Zheng:2014:BAbE:JPDC} is based on ELL and JAD. The input matrix is split into slices of $32$ rows (warp size). A zero padding is applied if the total rows in the input matrix are not multiple of $32$. The rows inside each slice are sorted in descending order of $\npr$ after which the nonzeros in each row are shifted to the left as in JAD and ELL. The columns inside the slices are then divided into $6$ groups. Each of these groups are stored in the ELL format. The bisection of the slices into groups improves load balancing and hence the performance of SpMV. One warp is assigned to one slice of BiELL. Each row of a group will be assigned a thread from the warp. These threads iterate over the other groups in the same slice in each iteration. Once completed, it is reassigned to the other slices. Each thread computes the partial result and stores in the shared memory. The final result is obtained with the help of reduction. The authors also propose Bisection JAD based on a similar technique.
Other schemes in this category include~\cite{Su:2012:c}.

\item[Sorted.]\label{sec:gpu:stfo:structural:sorted}Weber et. al.~\cite{Weber:2012:EGDS:J} extended CSR and proposed BIN-CSR to solve finite elements based sparse linear systems. It partitions the nonzeros into bins (a group of rows with a similar number of nonzeros). Each matrix row is processed using a thread and each bin is processed by a warp. The number of rows in a bin equals the size of warp and the number of columns in the bin is dependent on the largest row in the bin. Hence, this scheme introduces zero-padding like ELL. They also propose BIN-BCSR, which is a blocked version of BIN-CSR.
Kreutzer \etal~\cite{Kreutzer:2014:AUSM} proposed SELL-C-$\sigma$ that could be used on Intel multicore processors, Intel MICs, and GPUs. It is based on sliced ELL combined with SIMD vectorization. The C (chunk) indicates the width of the SIMD registers in the x86 architectures. On GPUs it is the number of threads per warp. The parameter $\sigma$ is the {\em sorting scope}, which is the number of rows that will be sorted to reduce the overhead, reduce padding, and increase performance. The rows within a chunk are stored column-wise and the chunks are stored one after another. 
Anzt et al.~\cite{anzt2014implementing} extended SELL-C-$\sigma$ and proposed SELL-P (also called Padded Sliced ELLPACK), specifically for GPU architecture. They introduce zero-padding to satisfy the memory constraints of the GPU architectures. 
Ashari et al.~\cite{Ashari:2014:FSMM:SC} proposed Adaptive Compressed Sparse Row (ACSR) that reduces both space and time requirements. It considers the thread divergence and load balancing issues, and reduces zero-padding (extra computations) required by ELLPACK and similar schemes. The synchronization overhead imposed by reduction operations are also reduced as compared to COO. The rows of the sparse matrix are moved into bins based on $\npr$. The rows with nearly equal $\npr$ are binned together. Each Bin is assigned a specific kernel for execution. Each kernel is configured to efficiently execute SpMV on those bins. For the bins that contain rows with large $\npr$, a master kernel using dynamic parallelism is invoked.
Wong \etal~\cite{Wong:2015:Ansm:NME} proposed a variant of ELLPACK-R and padded JAD (JDS) sparse storage schemes, called ELL-WARP, and an SpMV algorithm for finite element problems. The matrix is row sorted by $\npr$. All rows greater than the size of a warp are divided and repacked as new rows. They propose two kernels, ELL WARP (K1) and ELL WARP v2 (K2), also known as WPK1 and WPK2, respectively. Both these kernels initially sort the rows by length. WPK1 then arranges rows into groups of warp size and applies zero-pads accordingly, and reorders the data within each warp in a column-major coalesced pattern. In WPK2, the rows are subdivided recursively beginning when the number of elements a thread should execute exceeds the prescribed threshold. The number of elements a thread should execute are continued to be subdivided until the number reaches below the threshold. This is done to efficiently process the larger rows, which is in contrast to WPK1.
Other schemes in this category include~\cite{Muhammed:2019:SANM:APP}.

\item[Hybrid.]\label{sec:gpu:stfo:structural:hybrid}
The Hybrid format (also called HYB) was the first sparse storage scheme specifically developed for GPUs, developed by Bell and Garland~\cite{BellGarland2008} of Nvidia to eliminate the issues with ELL, mainly the loss of performance due to varying $\npr$. HYB stores the matrix into two parts; one stored in the ELL format and the other as COO. Generally, ELL is three times faster than COO for matrices with rows greater than 4000. Hence, a $k$th column is added to the ELL part if at least one third part of the matrix rows contains $k$ nonzeros. The remaining nonzeros in the rows are stored using the COO format. The CUSP~\cite{Cusp} implementation of HYB computes $k$ using $\npr$ histogram. 
Monakov and Avetisyan~\cite{Alex:2009:IBSM} proposed two schemes based on hybrid storage structures: first, a hybrid of BCSR (Block CSR) and BCOO (Block COO); and, second, a hybrid of BCSR, BCOO, and ELL. In the first hybrid technique, the sparse matrix is divided into slices with each slice containing a fixed number of consecutive rows. Within each slice (strip) the blocks are formed and are stored in BCOO format. The block column indices and offsets are stored from top row strip in one word. For each strip, the index of the first block in the strip is stored. Each of this strip is assigned a thread block. In order to reduce the zero elements within each block, they use the second hybrid technique (with ELL). To reduce the effects of padded zeros in ELL the rows with similar $\npr$ are reordered into a single strip.
Cao \etal~\cite{Cao:2010:ISMm:ICCASM} proposed ELLPACK-RP that is a hybrid of ELLPACK-R and JAD. ELLPACK-R has load imbalance problems due to varying $\npr$ while JAD has poor coalesced memory access. ELLPACK-RP can be constructed from ELLPACK-R by permuting the rows in descending order. A permutation of the row length array of ELLPACK-R is also required. A separate array is used for the storage of the permutations. They used one thread for two rows.
Kreutzer et al.~\cite{Kreutzer:2012:SMMo:IPDPSW}, in order to reduce the space overhead of extra padded zeroes in ELL, proposed the pJDS scheme that is based on ideas similar to ELLPACK, JDS, and sliced ELLPACK. The sparse matrix is shifted left after removing the zeroes as in ELL. Moreover, the rows are sorted based on the descending order of $\npr$. The sorted rows are then arranged into blocks, and each block is assigned to a CUDA block so that the load is balanced among the warps. Its major disadvantage is that the multiplication is performed by permutation, which could reduce the density of blocks and diagonal elements, prevent cache reuse, and degrade the performance.
Similar to HYB (hybrid of ELLPACK and COO), Liu \etal~\cite{liu2012sparse} proposed HEC, a hybrid of ELLPACK and CSR. HEC is designed considering the ease of using ILU-like preconditioners for solving sparse linear equation systems. For partitioning, the row length of ELL is found out by solving a minimum problem. To solve the minimum problem relative performance of both ELL and CSR is required, which requires some pretests. 
Maggioni \etal~\cite{Maggioni:2013:GSSo:IPDPSW} proposed a scheme that is hybrid of ELL and DIA (they proposed two schemes, one has been discussed earlier). The dense diagonals are stored as DIA (with padding if required) and the rest of the matrix is stored in ELL format. It is designed to improve the performance of Jacobi iterative solvers because the dense diagonals in form of DIA would improve the performance due to a better access of diagonal nonzeros. However, matrices with sparse diagonals render a lower performance compared to ELL and HYB.
Yang \etal~\cite{Yang:2013:Ooqm} proposed HDC, a hybrid of DIA and CSR addressing their respective limitations of inefficient storage of irregular diagonal matrices and imbalance in storing nonzeros. A threshold value for the number of nonzeros in a diagonal determines whether the diagonal is stored using DIA or CSR. 
Feng et al.~\cite{Feng:2012:Assm:CPE} proposed SHEC, a hybrid of Segmented Hybrid ELL (SHE: proposed in this same paper) and CSR to improve throughput and reduce memory usage on Nvidia GPUs. SHEC uses interleaved combination of matrix rows to improve coalesced memory access and balance the load among the threads. The input matrix is initially shifted left similar to ELL by removing zeros and a fixed number of rows ($cr$) are combined by an interleaved combination to form a new SHEC row. The value of $cr$ should be a power of 2 but will not exceed 32. The technique reduces to CSR and ELL when $cr=1$ and $cr=32$, respectively. The matrix is sorted and six segments are created based on $\npr$ to improve load balancing. The GPU kernel is similar to the CSR vector format where one warp is assigned to the $cr$ rows or one SHEC row.
Yang et al.~\cite{Yang:2018:Apcm:JCSS} proposed a hybrid storage scheme and partitioning scheme in which the matrix is partitioned into CSR and ELL formats.
Other schemes (\eg~\cite{Xiao:2021:CACa:TPDS,Yan:2014:Mboo:S,Bylina:2014:AERo}) exists based on hybrid storage of matrices. These are mentioned in other sections because their primary focus or storage method relate to other categories.

\end{description}
\subsubsection{Speed-Enhancement Techniques}
\label{sec:gpu:stfo:se}
All the scheme related to speed-enhancement
are primarily related to a storage scheme and hence have been discussed in the previous section (\ref{sec:gpu:stfo:structural}). Therefore, we mention the relevant schemes in the respective sections below. 

\begin{description}[leftmargin=0pt, itemsep=0.5em, topsep=0.5pt]
\item[Coalesced-Memory.]\label{sec:gpu:stfo:se.cm} The schemes in this category include~\cite{BellGarland2008, OberhuberSuzukiVacata2010,Vazquez:2010:Anaf:CPE,Vazquez:2012:Atot:PARCO,Sun:2011:OSfD:ICPP,Yang:2012:Aism:IPEMC,Feng:2011:OoSM:ICPADS,Yang:2011:Fsmm,Godwin:2012:Hsmm,Ashari:2014:Aetb,liu2015csr5,Dziekonski:2011:AMEA:PIER,Dang:2012:TSCF:PROCS,anzt2014implementing,Wong:2015:Ansm:NME, Cao:2010:ISMm:ICCASM, Kreutzer:2012:SMMo:IPDPSW, liu2012sparse, Feng:2012:Assm:CPE, YanLiZhangEtAl2014,Zhang:2016:ACSF, Maggioni:2013:AATf:PROCS, Maggioni:2013:AAAW:ICPP, Xiao:2021:CACa:TPDS}.

\item[Compression.] \label{sec:gpu:stfo:se.c}
The schemes in this category include~\cite{Wieczorek:2014:AEWo, RPR12, Tang:2013:Asmm,Tang:2015:AFoB:TPDS, Maggioni:2014:CAaC:IPDPSW, Bylina:2014:AERo}.

\item[Data-Reuse.] \label{sec:gpu:stfo:se.dr}
The schemes in this category include~\cite{Maggioni:2013:AAAW:ICPP}.

\item[Load-balanced.] \label{sec:gpu:stfo:se.lb}
The schemes in this category include~\cite{Buatois:2009:Cnca, BellGarland2008, Sun:2011:OSfD:ICPP,Yang:2012:Aism:IPEMC,Choi:2010:Maos,Choi:2010:Maos,liu2015csr5,Alex:2010:ATSM,Dziekonski:2011:AMEA:PIER,Dang:2012:TSCF:PROCS,Maggioni:2013:GSSo:IPDPSW,barbieri2015three,Zheng:2014:BAbE:JPDC,Weber:2012:EGDS:J,Kreutzer:2014:AUSM,anzt2014implementing,Ashari:2014:FSMM:SC,Wong:2015:Ansm:NME,Alex:2009:IBSM, Cao:2010:ISMm:ICCASM, Yang:2013:Ooqm, Feng:2012:Assm:CPE, Yang:2018:Apcm:JCSS, Xiao:2021:CACa:TPDS, Yan:2014:Mboo:S, Matam:2011:ASMV:ICPP, Muhammed:2019:SANM:APP}.

\item[Thread-Divergence.]\label{sec:gpu:stfo:se.td}
The schemes in this category include~\cite{OberhuberSuzukiVacata2010,Vazquez:2010:Anaf:CPE}.

\end{description}

\subsubsection{Compression-based Techniques}
\label{sec:gpu:stfo:cb}
This section reviews compression-based techniques for GPUs; see Section~\ref{sec:taxonomy:stfo} for definitions.
\begin{description}[leftmargin=0pt, itemsep=0.5em, topsep=0.5pt]
\item[Bandwidth-Specific.]\label{sec:gpu:stfo:cb.bs}
Tang et al.~\cite{Tang:2013:Asmm,Tang:2015:AFoB:TPDS} proposed a family of bit representation-optimized (BRO) sparse storage schemes on GPUs, including BRO-CSR, BRO-ELL, BRO-HYB, BRO-HYBR, and BRO-HYBR(S), and claim better performance compared to the original schemes (BRO-CSR over CSR, etc.). They compress the nonzero index data to reduce the memory bandwidth usage and improve SpMV on GPUs. The authors use lossless compression to reduce the amount of storage required by the matrices and hence reduce the memory bandwidth. The main idea of these techniques is to use delta encoding and represent the indices using the minimum number of required bits. A challenge of BRO would be to match the decompression on the GPU with its warp execution model.
Maggioni and Berger-Wolf~\cite{Maggioni:2014:CAaC:IPDPSW} extended their earlier work on Adaptive ELL (AdELL; see Section~\ref{sec:gpu:stfo:config}) and proposed the Compressed Adaptive ELL (CoAdELL) scheme, which exploits compression of indices using delta encoding and warp granularity. Differential encoding is used between contagious nonzeros so that the column width can be represented with a smaller number of bits. The major difference between AdELL and CoAdELL is the compression of column indices. Instead of directly storing the column indices, delta indices are stored after differential encoding. 
Yan et al.~\cite{Yan:2014:Mboo:S} extended HYB to propose HYB-R (HYB-recursive) with the aim to optimize the memory bandwidth usage of GPUs. The authors note that, for some matrices, the number of nonzeros in the COO part of HYB is higher than that in the ELL part of the HYB. HYB-R recursively partitions the matrix into COO and ELL. This results in a higher number of nonzeros in the ELL part compared to HYB.
Initially, they create two parts as before, the COO part and the ELL part. Then nonzeros are recursively added to the ELL part. Each new recursion is stored as a new ELL part. Therefore, the ELL part consists of smaller ELL parts and the rest of the matrix will be stored in the COO part. The sparse matrix is sorted initially according to the number of nonzeros before the recursive addition to the ELL part. Since HYB format has a good performance on GPUs, increasing the nonzeros in the ELL part of HYB increases the performance.
Nagasaka \etal~\cite{Nagasaka:2016:AMBO:PROCS} proposed the Adaptive Multi-level Blocking (AMB) format, which reduces memory traffic in SpMV computations to improve performance. They also include a number of optimization techniques including division and blocking of the given matrix, compression of the column indices and improvement in the re-usability of input vector element in the cache. A configuration mechanism determines the best parameters for each matrix data by estimating the memory traffic and predicting the performance of a given SpMV computation.
Other schemes in this category include~\cite{Grewe:2011:Agat}.

\item[Device-Device.]\label{sec:gpu:stfo:cb.cd}
Wieczorek \etal~\cite{Wieczorek:2014:AEWo} proposed a compression scheme for Markov matrices which takes advantage of the redundant rows. Encoding the column indices deferentially with respect to the row index it is observed that many of the rows are same which leads to storing the rows in a dictionary of unique deferentially encoded rows. Moreover, it was observed that the index sequence has fractal-like similarities. The authors proposed a meta run length encoding algorithm to store these sequences. Decompression is performed symbol by symbol depending upon the required values for operation.
Neelima et al.~\cite{RPR12} extended their earlier work and proposed the Bit-Level Single Indexing (BLSI) format to reduce the host-to-device memory transfer overhead and the pre-processing time due to data compression. A single array of size $\nnz$ is used to store the indices. The column information is embedded into the bits of row indices similar to CMSR~\cite{Kwiatkowska:2002:OSoL, Mehmoodthesis04}.

\item[Memory-Specific.]\label{sec:gpu:stfo:cb.ms}
Neelima and Raghavendra~\cite{Neelima:2011:CCOS} proposed the Column only SPaRse Format (CSPR). The row and column indices are combined and represented as a single value as in row-major representation of the matrix. The decoding of the data structure into row and column takes extra computation and time. A thread is assigned for each element in the matrix so as to increase the contiguous memory access. CSPR only performs well if the matrix has a large number of rows with small $\npr$.
Bylina \etal~\cite{Bylina:2014:AERo} proposed HYBIV (HYB with Indexed Value), a modification of HYB, in order to reduce the memory required for a transition rate matrix of Markov models on GPUs. All the unique elements in the Markov matrix are stored in the GPU read-only memory. The row and columns indices are stored in separate arrays, like the HYB and COO. HYBIV requires half the memory compared to HYB but it is limited to Markov matrices.
Yan \etal~\cite{YanLiZhangEtAl2014,Zhang:2016:ACSF} proposed the YASpMV framework in which they extended COO and introduced the Blocked Compressed COO (BCCOO) and Block Compressed COO plus (BCCOO+) schemes. In BCOO, a matrix is divided into blocks and each nonzero block is stored consecutively as COO. The BCCOO format uses bit flags to compress the row indices that eliminates condition check during the reduction of the partial sums. In BCCOO+, the sparse matrix is initially sliced vertically and the slices are aligned in top-down order. BCCOO is then applied on this newly obtained sliced matrix.
Other schemes in this category include~\cite{Yan:2014:Mboo:S}.

\end{description}

\subsubsection{Configurable Techniques}\label{sec:gpu:stfo:config}

Configurable techniques for GPUs (see Section~\ref{sec:taxonomy:stfo} for definition) are reviewed in this section. 

\begin{description}[leftmargin=0pt, itemsep=0.5em, topsep=0.5pt]
\item[Single-Format.]\label{sec:gpu:stfo:config.sf}
The schemes in this category include~\cite{Ashari:2014:Aetb, liu2015csr5, Dang:2012:TSCF:PROCS, Nagasaka:2016:AMBO:PROCS, Choi:2010:Maos,Choi:2010:Maos} and these are discussed in other sections of their primary categories.

\item[Multi-Format.]\label{sec:gpu:stfo:config.mf}
Matam and Kothapalli~\cite{Matam:2011:ASMV:ICPP} proposed an algorithm that analyzes the structure of a matrix and configures a suitable storage format for the matrix. A data structure is used which is a hybrid of CSR and ELL. It tries to balance the load among the threads automatically by tuning certain parameters.  
Other schemes in this category include~\cite{BellGarland2008,Alex:2009:IBSM, Buatois:2009:Cnca, Feng:2012:Assm:CPE, Yang:2018:Apcm:JCSS, Xiao:2021:CACa:TPDS, Yan:2014:Mboo:S, YanLiZhangEtAl2014,Zhang:2016:ACSF, Maggioni:2013:GSSo:IPDPSW, Yang:2013:Ooqm}

\end{description}
\subsection{Computations-focused (Cofo) Techniques (GPUs)}\label{sec:gpu:cofo}
 A number of works proposed SpMV algorithmic improvements on GPUs as discussed in the following section.
\subsubsection{Device Architectural}
\label{sec:gpu:cofo.da}
\begin{description}[leftmargin=0pt, itemsep=0.5em, topsep=0.5pt]

\item[Atomic Operations.]\label{sec:gpu:cofo:da.ao}
The schemes in this category with auxiliary (not primary) classifications include~\cite{Mukunoki:2013:OoSM, Grewe:2011:Agat, Liu:2015:LFCs:ASAP}.

\item[Cache-Specific.]\label{sec:gpu:cofo:da.cs}
Mukunoki and Takahashi~\cite{Mukunoki:2013:OoSM} extended their earlier work~\cite{Yoshizawa:2012:AToS:ICCSE} by exploiting the read-only data cache and shuffle instructions introduced by the latest GPU architectures at that time.
The other schemes in this category include~\cite{Dehnavi:2010:FSMV:TMAG,Xu:2012:OSMV:SNPD, Grewe:2011:Agat}.

\item[Compiler-Specific.]\label{sec:gpu:cofo:da.cos}
Baskaran and Bordawekar~\cite{baskaranoptimizing} were one of the first researchers to optimize and auto-tune the SpMV for GPUs. They developed compile time and run time strategies to improve the performance of SpMV on GPUs. 
Grewe and Lokhmotov~\cite{Grewe:2011:Agat} developed a special compiler that generates low level code for the high-level sparse matrix representation for GPUs. They also presented an optimizer that optimizes the generated parallel code. It optimizes the memory bandwidth, performs automatic vectorization, uses texture memory, caches the data, optimizes the reduction phase, and performs loop unrolling. 
The other schemes in this category include~\cite{Muhammed:2019:SANM:APP}.

\item[Dynamic Parallelism.]\label{sec:gpu:cofo:da.dp}
The/Other schemes in this category include~\cite{Ashari:2014:FSMM:SC,Muhammed:2019:SANM:APP}.

\item[Streams.]\label{sec:gpu:cofo:da.s}
The/Other schemes in this category include~\cite{Muhammed:2019:SANM:APP, Grewe:2011:Agat, Greathouse:2014:ESMM:SC}.

\item[Warp.]\label{sec:gpu:cofo:da.w}
Wijs and Bosnacki~\cite{Wijs:2012:IGSM} focused on Markov models and used indexed MSR  scheme (see \cite{Mehmood:2004:SDAo,UCAM-CL-TR-650})) such that the threads that are in the same warp access continuous memory. To access continuous memory in the GPU, they propose enumerating the non-zero elements of the sparse matrix using breadth-first search (BFS). They assign indices to the nonzeros of the matrix using a heuristics based on BFS. In one of the scheme, they use a single thread for a single row whereas in the second scheme they use two threads for a single row. The matrix is segmented into various partitions such that each segment consists of rows equal to the warp size. The other technique proposed by the authors implements half warp for a row.
Other schemes in this category include~\cite{Maggioni:2013:GSSo:IPDPSW,barbieri2015three, Kreutzer:2014:AUSM,Wong:2015:Ansm:NME, Feng:2012:Assm:CPE, Flegar:2017:OLIf, Maggioni:2013:AATf:PROCS, Maggioni:2013:AAAW:ICPP, Maggioni:2016:Otfs:JPDC, Liu:2015:LFCs:ASAP}.
\end{description}

\subsubsection{Algorithmic}\label{sec:gpu:cofo:algo}
\begin{description}[leftmargin=0pt, itemsep=0.5em, topsep=0.5pt]
\item[Auto-Tuning.]\label{sec:gpu:cofo:algo.at}
Guo and Wang~\cite{Guo:2010:ACPf:ICCIS} propose a configuration framework that can analyze a sparse matrix and automatically select parameters for CUDA kernel for SpMV to optimize the performance on specific GPUs. The authors investigate the effect of the number of threads, block size, and warp size on the performance of SpMV kernel. They report the best thread size to be between $Max~threads/2$ and $Max~threads$, where $Max~threads$ are the maximum number of threads supported by the GPU. The block size selection is made depending upon the model of GPU, and the warp size is selected to be 16 threads per warp. The matrix is converted to CSR and the framework identifies the best possible parameters and executes the kernel with these parameters.
Su and Keutzer~\cite{Su:2012:c} propose the Cocktail Format that takes advantages of different matrix formats by combining them. It partitions the input sparse matrix into several submatrices, each specialized for a given matrix format, with an auto-tuner. The considered schemes include DIA, BDIA, ELL,SELL, CSR, COO, BELL, SBELL, and BCSR.
Maggioni and Berger-Wolf~\cite{Maggioni:2013:AATf:PROCS} proposed Warp-Grained-ELL, an architectural optimization based on a heuristics, capable of reducing the cache memory access within a warp, to improve the performance of SpMV kernels. They also proposed an improvement to Sliced ELL~\cite{Alex:2010:ATSM}, 
based on warp granularity and reordering. The technique aims to improve the coalesced memory access and reduce the cache requests by changing the order in which the threads compute the nonzeros in the array. Sliced ELL~\cite{Alex:2010:ATSM} assumes that each slice will be assigned to a CUDA block of threads for execution. To avoid under utilization of GPU streaming multiprocessors (SMs) the block size and the warp size are decoupled from each other. The slice size which is equal to the warp size produces the best performance by achieving a better occupancy and data-structure efficiency.
Maggioni and Berger-Wolf~\cite{Maggioni:2013:AAAW:ICPP} proposed Adaptive ELL (AdELL), an extension of ELLPACK~\cite{GrimesKincaidYoung1979}. AdELL distributes variable number of threads to the rows depending upon the computational load, providing adaptive balancing of the warps for SpMV in GPUs. They proposed a new heuristic for assigning workload to threads. AdELL adapts itself to the structure of the input matrix. The proposed data structure is similar to Sliced ELL~\cite{Alex:2010:ATSM}. An atomic operation is performed for row reductions. The authors also discuss a heuristic for load assignment to the warps and another one for loop unrolling.
Maggioni and Berger-Wolf extended their work further and proposed AdELL+~\cite{Maggioni:2016:Otfs:JPDC}, an extension of their scheme CoAdELL, proposed earlier in~\cite{Maggioni:2014:CAaC:IPDPSW}. This technique implements adaptive warp balancing heuristics that reduces the preprocessing cost compared to AdELL~\cite{Maggioni:2013:AAAW:ICPP} . It also integrates a new auto-tuning approach which performs better than their previous scheme.
Greathouse and Daga~\cite{Greathouse:2014:ESMM:SC} proposed CSR-Stream (based on CSR) for SpMV on AMD based GPUs. The number of nonzeros to be processed by the wavefronts (warps) is fixed and the values are streamed into local scratchpad memory. The efficiency of this technique reduces with increasing $\npr$. Moreover, it becomes ineffective if the size of $\npr$ is more than the scratchpad memory size. To overcome this they dynamically determine whether a set of rows will be executed by Vector CSR (VCSR) or Scalar CSR (SCSR) (they refer to it as CSR-Adaptive).
Xiao et al.~\cite{Xiao:2021:CACa:TPDS} proposed the CASpMV framework (Customized and Accelerative framework for SpMV) specific to the Sunway processor architectures with the aim to improve storage limitations, load imbalance, and irregular memory accesses. CASpMV utilizes an auto-tuning four-way partitioning scheme for SpMV based on a statistical model of matrix sparsity to optimise for the computing architecture and memory hierarchy of Sunway processors. 
The other schemes in this category include~\cite{Choi:2010:Maos,Choi:2010:Maos,Dang:2012:TSCF:PROCS,Ashari:2014:FSMM:SC, Nagasaka:2016:AMBO:PROCS, Ashari:2014:Aetb, liu2015csr5, Alex:2010:ATSM, YanLiZhangEtAl2014,Zhang:2016:ACSF, baskaranoptimizing, Tan:2018:DaIo, Hou:2017:ASfP:IPDPSW, Elafrou:2019:B, Li:2013:S, Benatia:2019:Smpf}.

\item[Coalesced-Memory.]\label{sec:gpu:cofo:algo.cm}
The schemes in this category include~\cite{Wijs:2012:IGSM}.

\item[Data-Reuse.]\label{sec:gpu:cofo:algo.dr}
The schemes in this category include~\cite{Maggioni:2013:AATf:PROCS, Maggioni:2013:AAAW:ICPP}.

\item[Load-Balancing.]\label{sec:gpu:cofo:algo.lb}
Merril and Garland~\cite{Merrill:2016:Msmm} proposed Merge-based SpMV that reduces the imbalance of SpMV in CSR by splitting the total computations performed by the nested loops. They utilize parallel strategies originally developed for efficient merging of sorted sequences~\cite{Odeh:2012:MP-P:IPDPSW}. The main idea of this scheme is to depict the SpMV decomposition as a logical merger of the CSR row-offsets and CSR nonzero data, which is then split fairly among parallel SMs. This partition ensures that no single processing element would be burdened by long rows or a large number of zero-length rows.  Equal partitions from this logical merger are assigned to the processing elements such that, each processing element performs a two-dimensional search to isolate the corresponding region within each list that would comprise its share. The regions can then be treated as independent SpMV sub-problems and consumed directly from the CSR data structures using the CSR sequential technique.
Flegar and Anzt~\cite{Flegar:2017:OLIf} proposed a load-balanced GPU kernel based on COO for computing SpMV. The proposed kernel uses a ``warp vote function" along with a warp level segmented scan to avoid  atomic collisions. It also  uses  an ``oversubscribing" parameter to determine the number of threads allocated per each physical core. The three arrays in the COO scheme are sorted with respect to the row index to decrease the number of atomic operations such that the elements with the same row index require only one atomic operation.
Hou et al.~\cite{Hou:2017:ASfP:IPDPSW} proposed an auto-tuning framework for AMD APU platforms to find an appropriate binning scheme and select the appropriate kernel for each bin. The proposed binning method is a coarse grained method with different granularities. The auto-tuning technique selects the best granularity for a given matrix using machine learning techniques (C5.0). In their binning strategy, they combine multiple neighbouring rows to be a single virtual row. The granularity or the number of neighbouring rows selected is determined by the auto-tuner. The autotuner hence needs to be trained with appropriate data for the proper feature selection and prediction, which is non-trivial. Moreover, for irregular matrices with large $\nprv$, a fixed granularity will require each bin needing a different kernel. Executing a large number of kernels at the same time (which is subject to the device capability) requires multiple streams which is hardware dependent and hence might degrade the performance if the number of bins increases.
Mohammed et al.~\cite{Muhammed:2019:SANM:APP} proposed SURAA, a novel method for SpMV computations that groups matrix rows into different segments, and adaptively schedules various segments to different types of kernels. The segments of equal size are formed using the Freedman–Diaconis rule. The segments are assembled into three groups based on the $\anpr$ of the segments. For each group, multiple kernels are used to execute the group segments on different streams. Dynamic Parallelism is utilized to execute the group containing segments with the largest $\anpr$, providing improved load balancing and coalesced memory access, and hence more efficient SpMV computations on GPUs.
The schemes in this category include~\cite{Kreutzer:2014:AUSM, baskaranoptimizing, Greathouse:2014:ESMM:SC, Elafrou:2019:B, eguly:2012:Esmm:INPAR, Yoshizawa:2012:AToS:ICCSE, Ahamed:2012:Fsmm:HPCC, HellerOberhuber2012, Baxter13, Liu:2015:LFCs:ASAP}.

\item[Reduction.]\label{sec:gpu:cofo:algo.r}
The schemes in this category include~\cite{Liu:2015:Sssf:PARCO,liu2015csr5, Mukunoki:2013:OoSM, Grewe:2011:Agat, Flegar:2017:OLIf, Baxter13, Yin:2013:POfS}.

\item[Thread-Divergence.]\label{sec:gpu:cofo:algo.td}
Wang et al.~\cite{Wang:2010:Osmm:ICETC} proposed a few techniques to optimize the Vector CSR (VCSR) SpMV method and to avoid thread divergence. The threads are allocated to each row based on $\anpr$.
Other schemes include~\cite{Dehnavi:2010:FSMV:TMAG,Feng:2011:OoSM:ICPADS,Ashari:2014:FSMM:SC, Elafrou:2019:B, Baxter13} and are discussed in the categories of their primary focus.

\item[Thread-Mapping.]\label{sec:gpu:cofo:algo.tm}
Reguly and Giles~\cite{eguly:2012:Esmm:INPAR} proposed techniques to improve the CSR Vector (CSRV) kernel by assigning optimal number of threads per row in contrast to Bell and Garland~\cite{BellGarland2008} who use warp per row. The number of threads is $\{x | x=2^n\land x \geq \anpr, ~x,n \in \mathbb{Z^+}\}$. 
Yoshizawa and Takahashi~\cite{Yoshizawa:2012:AToS:ICCSE} proposed a kernel similar to Reguly and Giles~\cite{eguly:2012:Esmm:INPAR} except that the selection of threads per row is based on $\maxnpr$ instead of $\anpr$. 
Ahamed and Magoules~\cite{Ahamed:2012:Fsmm:HPCC} proposed improvements to finite element linear iterative solvers, specifically improvements to CUDA grid for CSR-based formats.
Verschoor and Jalba~\cite{Verschoor:2012:Aape:PARCO} studied the mapping of multiple block rows to thread rows while using BCSR~\cite{Buatois:2009:Cnca}. 
Heller and Oberhuber~\cite{HellerOberhuber2012} proposed Adaptive RgCSR, an extension of RgCSR~\cite{OberhuberSuzukiVacata2010}. The main disadvantage of RgCSR was that the threads per row were fixed (one thread per row), which resulted in the serialization of the computation due to variable $\npr$. Each group in Adaptive RgCSR is assigned a CUDA block and multiple threads are assigned to each row.
Baxter~\cite{Baxter13} developed the modernGPU library that provides a number of optimised primitives for linear algebra computing on GPUs. They proposed improvements to CSR kernels based on segmented reduction (see e.g.~\cite{Larsen:2017:Sfrs}). Initially, the sparse matrix nonzeros are multiplied to the vector elements in parallel and then a segmented reduction is performed to obtain the final vector. 
Yin et al.~\cite{Yin:2013:POfS} proposed improvements to CSR kernels. The input sparse matrix is divided into fragments (segments with varying number of rows). Each of these fragments is a multiple of the number of threads in the thread block and are assigned to a single block. The partial sum of each fragment is stored in shared-memory and later through reduction is written to the main memory.
Liu and Schmidt~\cite{Liu:2015:LFCs:ASAP} proposed LightSpMV, and optimisation of CSR kernels, which uses fine-grained dynamic allocation of matrix rows to threads. They proposed two approaches, one at the warp level and the other at the vector level. They use atomic operations and warp shuffle functions available in CUDA to implement this scheme.  
Abdelfattah \etal~\cite{Abdelfattah:2016:PooS:CPE} use dense matrix block operations in KBLAS library~\cite{Abdelfattah:2016:K} to compute dense matrix block products and optimise SpMV performance.
Other schemes include~\cite{Dehnavi:2010:FSMV:TMAG,Vazquez:2012:Atot:PARCO,Koza:2014:CMSF,Zheng:2014:BAbE:JPDC, Tang:2013:Asmm,Tang:2015:AFoB:TPDS, Guo:2010:ACPf:ICCIS, Muhammed:2019:SANM:APP, Wijs:2012:IGSM, Flegar:2017:OLIf, Maggioni:2013:AATf:PROCS, Maggioni:2013:AAAW:ICPP}.
\end{description}

\subsection{Auto-Selective Techniques (GPUs)}\label{sec:gpu:select}
This section reviews Auto-Selective Techniques for GPUs; see Section~\ref{sec:taxonomy:sefo} for definitions.

\subsubsection{Performance-Prediction (PP) Models}\label{sec:gpu:select.ppm}

\begin{description}[leftmargin=0pt, itemsep=0.5em, topsep=0.5pt]
\item[Deterministic.]\label{sec:gpu:select.ppm.d}
Zardoshti \etal~\cite{Zardoshti:2015:Asmr:S} proposed an adaptive scheme that runs tests on a sample of the matrices and selects the storage format for a given matrix during runtime. The authors do not specify the mechanism used for classification and identification of the parameters. Kubota and Takashi~\cite{Kubota:2011:OoSM} proposed a selection algorithm for SpMV on GPUs that analytically selects the best storage scheme based on $\npr$ and $\nprv$.
Guo et al.~\cite{Guo:2014:APMa:TPDS} proposed an analytical and profile-based PP model to predict SpMV execution times for COO, CSR, ELL, and HYB kernels, and auto-select the best kernel. They models incorporate device architectural features and matrix size but do not consider matrix sparsity structure. 
Guo and Wang~\cite{Guo:2014:Acpm:CPE} extend their earlier work (\cite{Guo:2014:APMa:TPDS}) to provide cross-architectural GPU models.
Other schemes in this category include~\cite{Su:2012:c}.

\item[Probabilistic.]\label{sec:gpu:select.ppm.p}
Li \etal~\cite{Li:2015:PAaO:TPDS} propose an optimization scheme based on calculating probability mass functions (PMF) to analyze the sparsity structure of matrices. They also propose probabilistic models for predicting the performance of COO, CSR, ELL, and HYB formats and then compute and compare the performance of the used formats for selecting the most suitable format. 

\end{description}

\subsubsection{Machine Learning}\label{sec:gpu:select.ml}

\begin{description}[leftmargin=0pt, itemsep=0.5em, topsep=0.5pt]
\item[Supervised.]\label{sec:gpu:select.ml.sup}
EL Zein and Rendell~\cite{Zein:2011:GoCs:CPE} propose a auto-selection method for CSR that selects either VCSR or SCSR kernel depending on $\npr$ using decision trees. 
Sedaghati \etal~\cite{Sedaghati:2015:ASoS} proposed a classification system based on decision trees using Simple Cart and BFTree implementations in the Weka library. They devised two feature sets for the purpose. 
Benatia \etal~\cite{Benatia:2016:SMFS:ICPP} proposed an auto-selection method for SpMV using the libSVM library for training and classification of the best storage scheme for a given matrix.
They extended their work in~\cite{Benatia:2018:B} where they used a weighted SVM with the help of pairwise classification approach (instead of multi-class SVM). The features used are similar to their previous work but in this case the feature combinations are selected based on the pair of storage formats for creating each pairwise model. 
Israt et al.~\cite{Nisa:2018:EMLB:IPDPSW} extended their earlier work (\cite{Sedaghati:2015:ASoS}) and compared the performance of the machine learning techniques, specifically the SVM and decision-tree based models, discussed by Benatia et al.~\cite{Benatia:2016:SMFS:ICPP} and Sedaghati \etal~\cite{Sedaghati:2015:ASoS}, along with multi-layer perceptrons (MLP), gradient boosting based models (XG-Boost) and ensembles of MLP for the prediction of the best sparse storage format for the SpMV computations of a matrix. 
Tan et al.~\cite{Tan:2018:DaIo} propose an auto-tuning SpMV library called SMATER that creates a learning model using data mining and analytical models. They use the C5.0 decision tree algorithm to generate a rule-set pattern with confidence data. A scorecard is used to select the best performing kernel for the predicted storage format.
Elafrou\,\etal~\cite{Elafrou:2019:B} discuss an auto-tuner named BASMAT for bottleneck-Aware SpMV on GPUs with low overhead and execution time based on CSR. The three bottlenecks considered in their work are memory bandwidth, memory latency, and thread imbalance. The optimizations considered for the bottlenecks include symmetry compression, unique value compression~\cite{Kourtis:2008:Osmm}, delta coding of indices,  vertical column blocking, and merge-spmv~\cite{Merrill:2016:Msmm}. They train a decision tree based classifier to predict one of the three considered performance bottlenecks and an appropriate optimization is selected after the prediction.

\end{description}

\subsubsection{Deep Learning}\label{sec:gpu:select.dl}

\begin{description}[leftmargin=0pt, itemsep=0.5em, topsep=0.5pt]
\item[Supervised.]\label{sec:gpu:select.dl.sl}
Muhammed \etal~\cite{Moh17, Mohammed:2020:DAnd:S} proposed DIESEL, a deep learning-based tool that predicts and executes the best performing SpMV kernel for a given matrix using using feature set developed by them. This is the only deep learning-based SpMV tool that we have found to date in the literature. They also proposed a range of new metrics and methods for performance analysis, visualization and comparison of SpMV tools.
\end{description}

\begin{footnotesize}
\begin{longtable}[t!]{@{}llp{1.9cm}p{1.9cm}p{1.9cm}P{0.6cm}@{}}
\caption[Classifications of SpMV Techniques (GPUs)]{Classifications of SpMV Techniques (GPUs)}\label{tab:gpu:summ}\footnotesize\\
    \toprule
\multirow{2}{*}{Name} & \multicolumn{4}{c}{Taxonomy} &  \multirow{2}{*}{SpMV}\\ 
    \cmidrule(lr){2-5} & Class & Primary & Secondary&Tertiary\\
    \midrule \endfirsthead
    \caption[Classifications of SpMV Techniques for (GPUs)] {Classifications of SpMV Techniques for (GPUs)}
    \tabularnewline
    \toprule
    \multirow{2}{*}{Name} & \multicolumn{4}{c}{Taxonomy} &  \multirow{2}{*}{SpMV}\\ 
    \cmidrule(lr){2-5} & Class &Primary & Secondary& Tertiary\\
    \midrule \endhead
    \bottomrule
    \multicolumn{6}{r}{\textit{Continued}}\\
    \endfoot
    \bottomrule
    \endlastfoot
    ACSR~\cite{Ashari:2014:FSMM:SC}  & \stfo & Sorted&\cofotd&Dynamic-Par& \pmb{\checkmark}\\
    AdELL~\cite{Maggioni:2013:AAAW:ICPP} & \cofo & Auto-Tuning&\stfocm&\cofotm& \pmb{\checkmark}\\
    AdELL+~\cite{Maggioni:2016:Otfs:JPDC}& \cofo &Auto-Tuning&Warp&--& \pmb{\checkmark}\\
    Ahamed \etal~\cite{Ahamed:2012:Fsmm:HPCC} & \cofo  &\cofotm&\cofolb&--&\pmb{\checkmark}\\
    AMB~\cite{Nagasaka:2016:AMBO:PROCS} & \stfo &Bandwidth&Single-Format&Auto-Tuning& \pmb{\checkmark}\\
    ArgCSR~\cite{HellerOberhuber2012} & \cofo & \cofotm&\cofolb&--& \pmb{\checkmark}\\
    Baskaran \etal~\cite{baskaranoptimizing} &\cofo &Compiler&Auto-Tuning&\cofolb& \pmb{\checkmark}\\
BCCOO~\cite{YanLiZhangEtAl2014} & \stfo &\mem& Multi-Format&Sliced& \pmb{\checkmark}\\
    BCCOO+~\cite{YanLiZhangEtAl2014} &\stfo &\mem& Multi-Format&Sliced& \pmb{\checkmark}\\
    BCSR~\cite{Buatois:2009:Cnca} & \stfo & Blocked& Single-Format& \stfolb &  \pmb{\checkmark}\\
    BELLPACK~\cite{Choi:2010:Maos} & \stfo &Blocked&Auto-tuning& Single-Format& \pmb{\checkmark}\\
    Benatia \etal~\cite{Benatia:2016:SMFS:ICPP} & \auto & \mls &--&--&\pmb{\checkmark}\\
    Benatia\,\etal~\cite{Benatia:2019:Smpf}& \cofo & Auto-Tuning &\stfolb& Sliced&\pmb{\checkmark}\\
    BestSF \etal~\cite{Benatia:2018:B}&\auto & \mls &--&--&\pmb{\checkmark}\\
    BiELL~\cite{Zheng:2014:BAbE:JPDC} & \stfo & Bisected&\cofotm&\stfolb& \pmb{\checkmark}\\
    BiJAD~\cite{Zheng:2014:BAbE:JPDC} & \stfo & Bisected &Thread Mapping&Load Balanced& \pmb{\checkmark}\\
    BIN-BCSR~\cite{Weber:2012:EGDS:J}& \stfo &Sorted&\stfolb&--& \pmb{\checkmark}\\
    BIN-CSR~\cite{Weber:2012:EGDS:J}& \stfo  &Sorted&\stfolb&Blocked& \pmb{\checkmark}\\
    BLSI~\cite{RPR12} & \stfo &Device-Device&Compression&--& \pmb{\checkmark}\\
    BRC~\cite{Ashari:2014:Aetb} &  \stfo &Blocked&Multi-Format&Auto-tuning& \pmb{\checkmark}\\
    BRO-x~\cite{Tang:2015:AFoB:TPDS} & \stfo &Bandwidth& Compression &--& \pmb{\checkmark}\\
    Cocktail~\cite{Su:2012:c} & \cofo & Auto-tuning&Deterministic& Bisected&\pmb{\checkmark}\\
    BTJDS~\cite{Abu-Sufah:2012:AEAf:HPCC} & \stfo & Blocked&--&--& \pmb{\checkmark}\\
    CASpMV~\cite{Xiao:2021:CACa:TPDS}& \cofo & Auto-Tuning &\stfolb&Multi-Format& \pmb{\checkmark}\\
    CDS~\cite{Godwin:2012:Hsmm} & \stfo & Blocked&\stfocm&--&\pmb{\checkmark}\\
    CMRS~\cite{Koza:2014:CMSF} & \stfo & Sliced&\cofotm&--& \pmb{\checkmark}\\
    CoAdELL~\cite{Maggioni:2014:CAaC:IPDPSW} & \stfo  &Bandwidth&Compression&--& \pmb{\checkmark}\\
    COO~\cite{BellGarland2008} & \stfo  &Flat&--&--& \pmb{\checkmark}\\
    CRSD~\cite{Sun:2011:OSfD:ICPP} & \stfo &Flat&\stfolb&\stfocm&\pmb{\checkmark}\\
    CSPR~\cite{Neelima:2011:CCOS} & \stfo &\mem&--&--& \pmb{\checkmark}\\
    CSR~\cite{BellGarland2008}& \stfo &Flat&--&--&\pmb{\checkmark}\\
    CSR5~\cite{liu2015csr5}& \stfo &Blocked&\stfolb&Single-Format&\pmb{\checkmark}\\
    CSR-Adaptive~\cite{Greathouse:2014:ESMM:SC} & \cofo &Auto-tuning&Streams&\cofolb& \pmb{\checkmark}\\
    CSR Vector~\cite{BellGarland2008, eguly:2012:Esmm:INPAR} & \cofo  & \cofotm&\cofolb&--&\pmb{\checkmark}\\
    DIA~\cite{BellGarland2008} & \stfo & \stfolb & \stfocm &--&\pmb{\checkmark}\\
    DIESEL~\cite{Moh17, Mohammed:2020:DAnd:S} & \auto & \dls &--&--&\pmb{\checkmark}\\
    eCSB~\cite{Tao:2014:Gasm:CPE} & \stfo & Blocked&Bandwidth&--& \pmb{\checkmark}\\
    Elafrou\,\etal~\cite{Elafrou:2019:B}& \auto &\mls&Auto-Tuning&\cofotd& \pmb{\checkmark}\\
    El Zein \etal~\cite{Zein:2011:GoCs:CPE} &\auto &\mls &--&--&\pmb{\checkmark}\\
    ELL~\cite{BellGarland2008} & \stfo & Flat &\stfolb&\stfocm& \pmb{\checkmark}\\
    ELL-R~\cite{Vazquez:2010:Anaf:CPE} & \stfo & Flat &\stfocm&\stfotd& \pmb{\checkmark}\\
    ELLPACK-RP~\cite{Cao:2010:ISMm:ICCASM} & \stfo & Hybrid &\stfocm&\stfolb& \pmb{\checkmark}\\
    ELLR-T~\cite{Vazquez:2012:Atot:PARCO,Vazquez:2010:Anaf:CPE}& \stfo & Flat & \cofotm& \stfocm& \pmb{\checkmark}\\
    ELL-WARP~\cite{Wong:2015:Ansm:NME} & \stfo &Sorted& \stfolb & \stfocm & \pmb{\checkmark}\\
    Flegar and Anzt~\cite{Flegar:2017:OLIf}& \cofo &\cofolb&Warp&\cofor& \pmb{\checkmark}\\
    Grewe \etal ~\cite{Grewe:2011:Agat} & \cofo  &Compiler&Cache&Streams& \pmb{\checkmark}\\
    Guo \etal~\cite{Guo:2010:ACPf:ICCIS} & \cofo  &Auto-Tuning&\cofotm&--&\pmb{\checkmark}\\
    Guo \etal~\cite{Guo:2014:Acpm:CPE} & \auto & Analytical &--&--&\pmb{\checkmark}\\
    Guo \etal~\cite{Guo:2014:APMa:TPDS} & \auto & Deterministic &--&--&\pmb{\checkmark}\\
    HDC~\cite{Yang:2013:Ooqm} & \stfo &Hybrid & \stfolb& Multi-Format &\pmb{\checkmark}\\
    HDIA~\cite{barbieri2015three} & \stfo &Sliced&\stfolb&\stfocm&\pmb{\checkmark}\\
    HEC \etal~\cite{liu2012sparse}  & \stfo &Hybrid&\stfolb& --&\pmb{\checkmark}\\
    HELLPACK~\cite{barbieri2015three} & \stfo &Sliced&\stfolb&\stfocm&\pmb{\checkmark}\\
    Hou et al.~\cite{Hou:2017:ASfP:IPDPSW}& \cofo  &\cofolb&Auto-Tuning&Sliced&\pmb{\checkmark}\\
    HYB~\cite{BellGarland2008}& \stfo  &Hybrid&\stfolb&Multi-format& \pmb{\checkmark}\\
    HYBIV~\cite{Bylina:2014:AERo} & \stfo  &\mem&Hybrid&Compression& \pmb{\checkmark}\\
    HYB-R~\cite{Yan:2014:Mboo:S} & \stfo &Bandwidth&Memory&Multi-Format& \pmb{\checkmark}\\
    ICSR~\cite{Yang:2012:Aism:IPEMC} & \stfo & Flat & \stfocm &\stfolb& \pmb{\checkmark}\\
    Israt et al.~\cite{Nisa:2018:EMLB:IPDPSW}& \auto & \mls &--&--&\pmb{\checkmark}\\
    JAD~\cite{Li:2012:Gpil:S,Cevahir:2009:FCGw} & \stfo  & Flat& \stfocm &--&\pmb{\checkmark}\\
    KBLAS SpMV~\cite{Abdelfattah:2016:PooS:CPE}& \cofo  & \cofotm&\cofolb& Blocked&\pmb{\checkmark}\\
    Kubota \etal~\cite{Kubota:2011:OoSM} & \auto & Deterministic & -- &--& \pmb{\checkmark}\\
    Li \etal~\cite{Li:2015:PAaO:TPDS} & \auto & Probabilistic &--&--& \pmb{\checkmark}\\
    LightSpMV~\cite{Liu:2015:LFCs:ASAP} & \cofo &\cofotm&Atomic Op& Warp&\pmb{\checkmark}\\
    Liu\, \etal~\cite{Liu:2015:Sssf:PARCO}& \stfo &Flat&Reduction& -- & \pmb{\checkmark}\\
    Matam \etal~\cite{Matam:2011:ASMV:ICPP}&\stfo &Multi-Format&\stfolb& -- & \pmb{\checkmark}\\
    Merge SpMV~\cite{Merrill:2016:Msmm} &\cofo &\cofolb&Flat&--&\pmb{\checkmark}\\
    moderngpu~\cite{Baxter13}& \cofo & \cofotm&Reduction&\cofotd&\pmb{\checkmark}\\
    Monakov \etal~\cite{Alex:2009:IBSM} & \stfo  &Hybrid &\stfolb&Multi-Format& \pmb{\checkmark}\\
    Mukunoki \etal~\cite{Mukunoki:2013:OoSM} & \cofo & Cache&Atomic Operations&\cofor&\pmb{\checkmark}\\
    PCSR~\cite{Dehnavi:2010:FSMV:TMAG}& \stfo &Flat&\stfotd& Cache& \pmb{\checkmark}\\
    pJDS~\cite{Kreutzer:2012:SMMo:IPDPSW} & \stfo  & Hybrid &\stfolb&--& \pmb{\checkmark}\\
    RgCSR~\cite{OberhuberSuzukiVacata2010} & \stfo  & Flat&\stfocm& \stfotd& \pmb{\checkmark}\\
    SCOO~\cite{Dang:2012:TSCF:PROCS}  & \stfo  & {Sliced} &\stfocm & Auto-tuning& \pmb{\checkmark}\\
    Sedaghati \etal~\cite{Sedaghati:2015:ASoS} &  \auto & \mls &--&--&\pmb{\checkmark}\\
    SELL-C-$\sigma$~\cite{Kreutzer:2014:AUSM} & \stfo & Sorted &\cofolb & \stfolb & \pmb{\checkmark}\\
    SELL-P~\cite{anzt2014implementing} & \stfo  &  Sorted&\stfolb&\stfocm& \pmb{\checkmark}\\
    SELLPACK~\cite{Alex:2010:ATSM} & \stfo  & {Sliced} &\stfolb& Auto-Tuning& \pmb{\checkmark}\\
    SELLR-T~\cite{Dziekonski:2011:AMEA:PIER} & \stfo  & {Sliced} &\stfocm&\stfolb& \pmb{\checkmark}\\
    SHEC~\cite{Feng:2012:Assm:CPE} & \stfo  &Hybrid &\stfolb&Multi-Format& \pmb{\checkmark}\\
    SIC~\cite{Feng:2011:OoSM:ICPADS} &\stfo  & Sliced&\stfocm&\cofotd& \pmb{\checkmark}\\
SMATER~\cite{Tan:2018:DaIo}&\auto & \mls &--&--&\pmb{\checkmark}\\
    SURAA~\cite{Muhammed:2019:SANM:APP}& \cofo & \cofolb & Dynamic-Par&Streams&\pmb{\checkmark}\\
Verschoor \etal~\cite{Verschoor:2012:Aape:PARCO} &\cofo & \cofotm &--&--&\pmb{\checkmark}\\
    Wang \etal~\cite{Wang:2010:Osmm:ICETC} &\cofo  & \cofotd&\cofotm&\cofolb& \pmb{\checkmark}\\
    Maggioni~\etal \cite{Maggioni:2013:GSSo:IPDPSW} & \stfo  & Sliced&\stfolb&Warp&\pmb{\checkmark}\\
    Maggioni~\etal \cite{Maggioni:2013:GSSo:IPDPSW} & \stfo  & Hybrid&\stfolb&Multi-Format&\pmb{\checkmark}\\
    Warp-GELL~\cite{Maggioni:2013:AATf:PROCS} & \cofo  &Auto-Tuning&Warp&Data-Reuse& \pmb{\checkmark}\\
    Wieczorek \etal~\cite{Wieczorek:2014:AEWo} & \stfo  &Device-Device&Compression&--& \pmb{\checkmark}\\
    Wijs \etal~\cite{Wijs:2012:IGSM} & \cofo  & Warp &\cofocm& \cofotm& \pmb{\checkmark}\\
    WPK1/WPK2~\cite{Wong:2015:Ansm:NME}&  \stfo   & Sorted&\stfolb& Warp&  \pmb{\checkmark}\\
    Xu \etal~\cite{Xu:2012:OSMV:SNPD} &  \stfo  & Blocked&Cache&--&  \pmb{\checkmark}\\
    Yang~\etal \etal~\cite{Yang:2011:Fsmm} & \stfo  & Blocked&\stfocm&--&  \pmb{\checkmark}\\
    Yang et al.~\cite{Yang:2018:Apcm:JCSS}&  \stfo  &Hybrid &\stfolb & Multi-Format&  \pmb{\checkmark}\\
    Yin \etal~\cite{Yin:2013:POfS} & \cofo &\cofotm& \cofor &--&\pmb{\checkmark}\\
    Yoshizawa \etal~\cite{Yoshizawa:2012:AToS:ICCSE} & \cofo  & \cofotm&\cofolb&--& \pmb{\checkmark}\\
    Zardoshti \etal~\cite{Zardoshti:2015:Asmr:S} & \auto & Deterministic&--&--& \pmb{\checkmark}\\
    \end{longtable}
\end{footnotesize}
\normalsize

\subsection{Summary and Analysis (GPUs)}
\label{sec:gpu:summary}

In this section, we have reviewed the various SpMV works proposed for the GPU architectures. 
The GPU techniques are very rich in terms of the novelty of approaches, and the breadth and depth of the proposals and evaluations. 
Table~\ref{tab:gpu:summ} gives a summary of the discussed GPU techniques. Column $1$ gives the SpMV scheme names in alphabetical order. Some schemes do not have a name and are represented by the author names. 
Column $2$ gives information whether the scheme is Stfo, Cofo, or Auto-Selective; 
see the taxonomy in Figure~\ref{fig:taxo}. 
Columns 3 to 5 give primary, secondary, and tertiary classifications of the schemes. Some schemes do not have all three classifications due to them addressing one or two performance issues and in these cases the respective entries do not have a classification. Some other schemes have addressed more than three performance issues but we have chosen to only include up to three in the table.     
Some performance issues belong to multiple L1 categories, 
such as Load-Balancing of an SpMV technique could be improved by storage structures (Stfo), algorithmically (Cofo), or both. To address this, we have used blue colour to represent Stfo categories, green colour for Cofo categories, and maroon colour if a scheme belongs to the supervised learning category of deep learning; the black colour is used if there is no confusion in category names. For example, the scheme ACSR is primarily a storage focussed scheme (\stfo) with ``Sorted'' data structure but its secondary category is ``Thread-Divergence'' (\cofotd) addressed by the authors by making algorithmic improvements (rather than by storage or data structure improvements). The scheme also exploits the Dynamic-Parallelism primitives offered by GPU but this is listed in text in black because there is not other category with the same name.          

The works focusing on Stfo (60) is higher than that of Cofo (28). Regarding Stfo techniques, larger number of works focus on ``Load-Balancing'' (31), ``Coalesced'' (19), ``Flat'' (14), and ``Sliced'' (12) techniques. Whereas, Cofo schemes focus on ``Thread-Mapping'' (21), ``Auto-tuning'' (17), and ``Load-Balancing'' (14). For the past five years an increase in ``Load-Balancing'' and ``Auto-tuning'' techniques is noticed, whereas the ``Thread-Mapping'' techniques have been in decline. Past three years have also seen the rise of ``Auto-Selective'' techniques (14). Later in Section~\ref{sec:future}, further discussions on the future research direction and open challenges are covered.

\section{SpMV Techniques for CPUs}\label{sec:cpu}

The SpMV techniques developed to be executed on CPUs are reviewed in this section. As mentioned earlier we do not intend to provide an extensive review of the CPU techniques, rather to develop a common taxonomy for all architectures, which could encourage developments of SpMV techniques for heterogeneous architectures.

\subsection{Storage-focused (Stfo) Techniques (CPU{s})}\label{sec:cpu:stfo}

\subsubsection{Structural Techniques}\label{sec:cpu:stfo:struct}
\begin{description}[leftmargin=0pt, itemsep=0.5em, topsep=0.5pt]
\item[Flat.]\label{sec:cpu:stfo:flat}
The earliest flat schemes devised for CPU architectures are 
COO~\cite{saad1994sparskit}, 
CSR~\cite{saad1994sparskit}, 
ELLPACK~\cite{GrimesKincaidYoung1979},  
JAD (or JDS)~\cite{Saad:1989:KSMo} 
and 
DIA~\cite{saad1994sparskit}.
The Modified Sparse Row (MSR) format~\cite{saad1994sparskit} differs from CSR in that it stores the matrix off-diagonal nonzeros separately, which speeds up the divide-by-diagonal operation in iterative methods. Modified Sparse Column (MSC) stores a matrix by columns.   
The Compact MSR (CMSR) format~\cite{Kwiatkowska:2002:OSoL, Mehmoodthesis04} takes advantage of the sparsity structure present in matrices arising from Markovian models (Markov Chains, Markov Decision Processes (MDPs), etc.) that have relatively fewer distinct nonzeros in the matrix. CMSR stores the distinct nonzeros in a separate array, and, to save space, stores both the nonzero index location as well as the nonzero column location in a single unsigned integer. Bitwise operations allow fast and parallel extraction of the required indices during SpMV execution.    
Ekambaram and Montagne~\cite{An:2003:AACS} extended JAD/JDS to propose Transposed JDS (TJDS or TJAD), which removed the need for the permutation array in JAD.
The Generalized Row/Column Storage (GCRS/GCCS) format~\cite{Shaikh:2015:Essf:HPCSIM} was proposed for matrices with dimensions greater than two. GCRS maps an $n+1$ dimensional array used to store an n-dimensional matrix to a $2D$ array using a mapping algorithm proposed by the authors.

\item[Blocked.]\label{sec:cpu:stfo:blocked} 
The Blocked Compact MSR (BCMSR) scheme~\cite{Mehmood2003-bdd, Mehmoodthesis04} is an extension of CMSR~\cite{Kwiatkowska:2002:OSoL, Mehmoodthesis04}. It divides the matrix into blocks and stores both the individual blocks and the high-level block structure of the matrix using the Compact MSR scheme. The aim was to parallelise solution of Markov models with high efficiency. BCMSR was later used to solve sparse linear equations systems on shared-memory machines~\cite{Kwiatkowska:Dpos:MASCOT} and distributed-memory systems~\cite{UCAM-CL-TR-650, Mehmood:APIM:MASCOTS, Zhang:AWPo:DSN}.  
Other blocked formats are 
BBCS~\cite{Stathis2004}, 
BCSR~\cite{Vuduc:2002:POaB:SC}, 
UBCSR~\cite{Vuduc:2005:Fsmm}, 
BCRS~\cite{Barret:1995:TftS,Stathis2004},
COOCOO~\cite{imecek:2012:SSMS:HPCC}, 
CSR2~\cite{Bian:2020:CANF:CCGRID}, 
ZAKI~\cite{Usman:2019:ZASM:S}, ZAKI+~\cite{Usman:2019:ZAML:ACCESS}, 
and CSB~\cite{Bulucc:2009:Psma}. 

\item[Sliced.]\label{sec:cpu:stfo:sliced}
Vuduc~\cite{Vuduc2003} proposed the Row Segmented Diagonal (RSDIAG) format, which is based on DIA. The scheme divides the matrix into row slices such that each slice consists of diagonal elements more than one or equal to the number of rows in a slice.
Other schemes include~\cite{Zhang:2018:VPSM, liu2015csr5}.

\item[Tree.]\label{sec:cpu:stfo:tree}
Abdali and Wiset~\cite{Abdali:1989:Ewqr} performed quantitative analysis on the suitability of the usage of the quadtrees for the representation of both sparse and dense matrices compared to the flat storage. They used the REDUCE~\cite{hearn1970reduce} implementation of quadtrees and flat representation of matrices and concluded that quadtrees provide speedups in matrix operations and inversion.
Dvorsky and Kratky proposed~\cite{dvorsky2004multi} a multi-dimensional scheme based on R-trees and BUB trees~\cite{fenk2002bub} where the space for storing the indices are higher than CSR but it has a better access rate to the matrix elements.
Balk \etal~\cite{Balk:2004:BBST} proposed a scheme using balanced binary search tree, the AVL Red-Black tree. Since these data structures have an access time of $O(\ln(n))$, the authors compared the time taken for this scheme against the traditional hash map memory allocation.
Simecek \etal~\cite{Simecek:2012:MQFf:SYNASC} extended their earlier work~\cite{Simecek:2009:SMCU:SYNASC} and proposed a Minimal QuadTree matrix storage format (MQT) and applied compression on it to reduce storage requirements. They do not use the matrix for SpMV computations, rather the purpose is to compactly store sparse matrices on disks.  
Zhang \etal~\cite{zhang2016efficient} proposed the COEQT scheme based on quadtrees and cache memory to compute SpMV. The authors recursively subdivide the sparse matrix into smaller sub-regions using quadtrees so that the sub-regions can be stored in the cache to improve the data locality.
Other tree formats include CMSR~\cite{Kwiatkowska:2002:OSoL, Mehmoodthesis04} and BCMSR~\cite{Mehmood2003-bdd, Mehmoodthesis04}.

\item[Hierarchical.]\label{sec:cpu:stfo:hiearchical}
Langr \etal~\cite{LangrSimecekTvrdikEtAl2012} proposed an Adaptive Hierarchical Blocking format (AHBSF) to store sparse matrices. They divide the matrix into blocks and store these blocks as a two level hierarchical data structure. The first level, level-0, is used to store the nnz blocks, which are stored as sub-matrices of the original sparse matrix and level-1 consists of a block matrix that stores pointers to the blocks in the first level. In the proposed technique for each block an efficient storage scheme is selected out of CSR, COO, dense and Bitmap and for the block matrix COO format is used. 
Other works include HiSM~\cite{Stathis2004,Stathis:AHsm:IPDPS},
CSRCOO~\cite{imecek:2012:SSMS:HPCC}, and BCMSR~\cite{Mehmood2003-bdd, Mehmoodthesis04, UCAM-CL-TR-650}.  

\item[Hybrid.]\label{sec:cpu:stfo:hybrid}
Yuan \etal~\cite{Yuan:2010:OSMV:HPCC} proposed two storage formats, DDD-Naïve and DDD-Split, based on a combination of DIA and CSR with the aim to improve DIA for matrices that have a smaller number of dense diagonals.
Other hybrid formats include AHBSF~\cite{LangrSimecekTvrdikEtAl2012}.

\end{description}

\subsubsection{Speed-Enhancement Techniques (CPUs)}\label{sec:cpu:stfo:se}
\begin{description}[leftmargin=0pt, itemsep=0.5em, topsep=0.5pt]

\item[Load-Balancing.]\label{sec:cpu:stfo:se.lb}
The schemes in this category include TJAD~\cite{An:2003:AACS}, ELL~\cite{GrimesKincaidYoung1979}, 
CMSR~\cite{Kwiatkowska:2002:OSoL, Mehmoodthesis04}, BCMSR~\cite{Mehmood2003-bdd, Mehmoodthesis04}, DIA~\cite{saad1994sparskit}, 
and 
MSR~\cite{saad1994sparskit}.

\item[Memory-Alignment.]\label{sec:cpu:stfo:se.ma}
The schemes in this category include TJAD~\cite{An:2003:AACS}, ELL~\cite{GrimesKincaidYoung1979}, JAD~\cite{Saad:1989:KSMo}, 
DIA~\cite{saad1994sparskit} 
and 
MSR~\cite{saad1994sparskit}.\end{description}

\subsubsection{Compression-based Techniques}\label{sec:cpu:stfo:cb}
\begin{description}[leftmargin=0pt, itemsep=0.5em, topsep=0.5pt]
\item[Bandwidth-Specific.]\label{sec:cpu:stfo:cb.bs}
The schemes in this category include 
SBSRS~\cite{ShahnazUsman2011}, AHBSF~\cite{LangrSimecekTvrdikEtAl2012},  
DDD-Naïve~\cite{Yuan:2010:OSMV:HPCC}, 
and 
DDD-Split~\cite{Yuan:2010:OSMV:HPCC}.

\item[Device-Device.]\label{sec:cpu:stfo:cb.cd}

Willcock and Lumsdaine~\cite{Willcock:2006:Asmc} proposed two storage formats based on lossless compression, Delta Coded Sparse Row (DCSR) and Row Pattern CSR (RPCSR). These schemes were designed for storage on disk to reduce the storage requirements and CPU-disk transfer times. 
Other works in this category include CMSR~\cite{Kwiatkowska:2002:OSoL, Mehmoodthesis04} and 
BCMSR~\cite{Mehmood2003-bdd, Mehmoodthesis04},  

\item[Memory-Specific.]
\label{sec:cpu:stfo:cb.ms}
Karakasis \etal~\cite{Karakasis:2013:AECF:TPDS} proposed an extension of CSR called Extended Compression format (CSX) to improve the performance of SpMVs on NUMA and SMP architectures.
Chen \etal~\cite{Chen:2018:AeSc:CPE} proposed a technique called CSR-SIMD to take full advantage of SIMD acceleration in the CPUs. The nonzero elements are compressed into variable length data with consecutive (coalesced) memory access. This improves the data locality for matrix and vector, and floating point operations are completely vectorized on the wider SIMD units.
Other schemes in this category include BBCS~\cite{Stathis2004}, HiSM~\cite{Stathis2004,Stathis:AHsm:IPDPS}, MQT~\cite{Simecek:2012:MQFf:SYNASC}, Balk \etal~\cite{Balk:2004:BBST}, BCRS~\cite{Barret:1995:TftS,Stathis2004}, CMSR~\cite{Kwiatkowska:2002:OSoL, Mehmoodthesis04}, BCMSR~\cite{Mehmood2003-bdd, Mehmoodthesis04}, and CSB~\cite{Bulucc:2009:Psma}.
\end{description}

\subsubsection{Configurable Techniques}\label{sec:cpu:stfo:ct}
\begin{description}[leftmargin=0pt, itemsep=0.5em, topsep=0.5pt]
\item[Single-Format.]\label{sec:cpu:stfo:ct.sf}
Vuduc and Moon~\cite{Vuduc:2005:Fsmm} extended classical BCSR and proposed UBCSR (Unaligned BCSR) to store sparse matrices with unaligned multiple structures. The matrix is split as the sum of sub-matrices and each sub-matrix is stored using the UBCSR format. The blocks in UBCSR can have variable size. 

\end{description}

\subsection{Computation-focused (Cofo) Techniques (CPUs)}\label{sec:cpu:cofo}

\subsubsection{Device Architectural}\label{sec:cpu:cofo.da}

Chen \etal~\cite{Chen:2020:tAtl:INS} propose a two-phase SpMV (tpSpMV) kernel for many-core and multi-core architectures with a focus on SW26010 CPUs. The two-phase strategy overcomes the computational scale limitations. Moreover, an adaptive partitioning strategy is proposed that utilizes a local memory caching technique for the two phases to reduce memory access latency. Data reduction, memory alignment, and pipeline techniques are utilized to improve the bandwidth usage and hence overall performance. 
Other schemes include COEQT~\cite{zhang2016efficient}. 

\begin{description}[leftmargin=0pt, itemsep=0.5em, topsep=0.5pt]
\item[Compiler-Specific.]\label{sec:cpu:cofo.da.compiler}
The schemes in this category include OSKI~\cite{Vuduc:2005:OAlo}, CSRLen~\cite{Aktemur:2018:Asmm:CPE}, and ~\cite{An:2016:AWPf:SC}. 

\item[Streams.]\label{sec:cpu:cofo.da.stream}
Guo and Gropp~\cite{Guo:2010:OSDS,Guo:2013:Aots} proposed the SpMV storage scheme Stream-CSR (S-CSR) based on CSR. It utilizes the pre-fetch data stream facility for increased memory bandwidth offered by IBM POWER architecture. It allows multiple computation groups to prefetch independent data streams without overlap so that the prefetch can be increased to accelerate the performance. 
\end{description}

\subsubsection{Algorithmic Techniques}\label{sec:cpu:cofo:algo}
\begin{description}[leftmargin=0pt, itemsep=0.5em, topsep=0.5pt]
\item[Auto-Tuning.]\label{sec:cpu:cofo:algo.at}
Many auto-tuning tools have been developed to improve the SpMV performance on CPUs. Most auto-tuning techniques use optimizations such as loop unrolling, explicit removal of unnecessary dependencies in code, and selection of optimized parameters. 
Vuduc \etal~\cite{Vuduc:2005:OAlo} proposed a tool named OSKI that combined both runtime and compile-time optimizations to auto-tune linear algebra kernels on CPUs. 
A similar tool, ATLAS, developed by Whaley and Dongarra~\cite{Whaley:1998:ATLA:SC} provides efficient implementations of various BLAS routines.
Im \etal~\cite{Im:2004:SOFf} introduced a framework named SPARSITY that auto-tunes sparse matrix operations by selecting different parameters such as block size using mathematical models. 
Venkat \etal~\cite{An:2016:AWPf:SC} proposed a compiler and runtime optimizer for parallelizing SpMVs with loop dependencies. The tool collects data dependence information and achieves wavefront parallelization. The proposed tool can parallelize Gauss-Seidel iterative solver and ILU0 method (see also~\cite{Zhang:AWPo:DSN} for wavefront parallelisation of Markov chains using BCMSR (also called modified MTBDDs)~\cite{Mehmood2003-bdd, Mehmood:2004:SDAo}). 
Vuduc \etal~\cite{Vuduc:2002:POaB:SC} proposed the Blocked CSR (BCSR) technique and heuristics for tuning the block parameters. 
Other works in this category include RSDIAG~\cite{Vuduc2003}, 
and 
SMAT ~\cite{Li:2013:S}.

\item[Data-Reuse.]\label{sec:cpu:cofo:algo.dr}
The schemes in this category include S-CSR~\cite{Alappat:2020:ARAC}.

\item[Load-Balancing.]\label{sec:cpu:cofo:algo.lb}
Catal{\'a}n \etal~\cite{S:2020:TaAa} parallelize the SpMV kernel in LASs (Linear Algebra routines on OmpSs) library. They improve the SpMV which is already based on OpenMP tasking and nesting by dynamically deciding the number of OpenMP tasks to be created. The dynamic allocation improves the resource utilization and load-balancing.
Bian \etal~\cite{Bian:2020:CANF:CCGRID} propose an extension of CSR named CSR2 suitable for SIMD architectures. The value and column arrays in CSR are divided by selecting  a block size and adding the required padding to fully utilize the SIMD vectorization function of the processor. 
Aktemur~\cite{Aktemur:2018:Asmm:CPE} proposed a variation of CSR storage scheme called CSRLen and associated SpMV implementation called CSRLenGoto. CSRLenGoto focus on inner loop unrolling and use of goto statements. 
Alappat:2020:ARAC \etal~\cite{Alappat:2020:ARAC} proposed a novel colouring algorithm and library called recursive algebraic coloring engine (RACE) to parallelize the SpMV multiplication operations of symmetric matrices. The proposed technique improves the load balancing and data redundancy and improves the performance of SpMV on CPUs.
Other schemes include COEQT~\cite{zhang2016efficient}, tpSpMV~\cite{Chen:2020:tAtl:INS},~\cite{Bian:2020:RoAt}, 
and~\cite{Zhang:2018:VPSM}, 

\item[Memory-Alignment.]\label{sec:cpu:cofo:algo.ma}
The schemes in this category include S-CSR~\cite{Guo:2010:OSDS,Guo:2013:Aots}, tpSpMV~\cite{Chen:2020:tAtl:INS}.

\item[Thread-Mapping.]\label{sec:cpu:cofo:algo.tm}
Bian \etal~\cite{Bian:2020:RoAt} utilize the AVX2 (Advanced Vector Extension 2) instruction sets for SIMD processors to enhance the performance of SpMV computations on CPUs.
Zhang \etal~\cite{Zhang:2018:VPSM} proposed a sliced ELLPACK format and an efficient SpMV kernel for the proposed format utilizing AVX, AVX2, and AVX512 instruction sets. The proposed technique is a part of the PETSc library.
The schemes in this category include CSR-SIMD~\cite{Chen:2018:AeSc:CPE}, ZAKI~\cite{Usman:2019:ZASM:S}, ZAKI+~\cite{Usman:2019:ZAML:ACCESS}.
\end{description}

\subsection{Auto-Selective Techniques (CPUs)}\label{sec:cpu:autos}

\begin{description}[leftmargin=0pt, itemsep=0.5em, topsep=0.5pt]
\item[Supervised.]\label{sec:cpu:select.ml.sl}
Li \etal~\cite{Li:2013:S} developed an SpMV Auto-Tuning System (SMAT) that accepts a sparse matrix in CSR and automatically determines the best matrix storage format (CSR, CSC, DIA, COO, and BSR) at runtime. SMAT uses a machine learning model developed using C5.0 decision trees, trained with matrix features.
Elaforu \etal~\cite{DBLP:journals/corr/ElafrouGK15} developed a selection-based tool using Naive Bayes classification for SPMV on CPUs. They did not select a format through classification rather selected optimizations for various matrix classes. Usman \etal~\cite{Usman:2019:ZASM:S} proposed the ZAKI tool for predicting the optimal number of processes for SpMV computations of an arbitrary sparse matrix on a distributed memory machine using decision trees, random forest, and gradient boosting, claiming this to be the first such work. The authors also discuss the potential application of their tool for optimization of energy efficiency of SpMV computations.
Usman \etal~\cite{Usman:2019:ZAML:ACCESS} extend their earlier work and proposed the ZAKI+ tool to predict the best SpMV parallelisation configuration in terms of the data distribution, the optimal number of processes, and mapping strategy on distributed memory machines. They claim this to be the first such work and that their tool ZAKI+ provides optimal process mapping better that the MPI default mapping policy.
\end{description}

\begin{footnotesize}
\begin{longtable}[t!]{@{}llp{2cm}p{2.3cm}p{2.3cm}P{0.6cm}@{}}
\caption[Classifications of SpMV Techniques (CPUs)]{Classifications of SpMV Techniques (CPUs)}\label{tabcpu.summ}
    \footnotesize\\
    \toprule
\multirow{2}{*}{Name} & \multicolumn{4}{c}{Taxonomy} &  \multirow{2}{*}{SpMV}\\
    \cmidrule(lr){2-5} & Class & Primary & Secondary&Tertiary\\
    \midrule \endfirsthead
    \caption[Classifications of SpMV Techniques (CPUs)] {Classifications of SpMV Techniques (CPUs)}
    \tabularnewline
    \toprule
    \multirow{2}{*}{Name} & \multicolumn{4}{c}{Taxonomy} &  \multirow{2}{*}{SpMV}\\
    \cmidrule(lr){2-5} & Class &Primary & Secondary& Tertiary\\
    \midrule \endhead
    \bottomrule
    \multicolumn{6}{r}{\textit{Continued}}\\
    \endfoot
    \bottomrule
    \endlastfoot
    Abdali and Wiset~\cite{Abdali:1989:Ewqr}& \stfo & Tree & -- & -- & --\\
    AHBSF~\cite{LangrSimecekTvrdikEtAl2012} & \stfo & \hierarchical & Hybrid & \bandw & --\\
    ATLAS~\cite{Whaley:1998:ATLA:SC}& \cofo & \cofoat & --  & -- & \pmb{\checkmark}\\
    Balk \etal~\cite{Balk:2004:BBST}& \stfo & Tree & \mem & -- & --\\
    BBCS~\cite{Stathis2004}& \stfo & Blocked & \mem & -- & \pmb{\checkmark}\\
    BCMSR~\cite{Mehmood2003-bdd, Mehmoodthesis04}& \stfo & Blocked & \mem & \cofolb & \pmb{\checkmark}\\
    BCRS~\cite{Barret:1995:TftS,Stathis2004}& \stfo & Blocked  & \mem  & --  & \pmb{\checkmark}\\
    BCSR~\cite{Vuduc:2002:POaB:SC}&\cofo & \cofoat & Blocked & --  & \pmb{\checkmark}\\
    Bian \etal~\cite{Bian:2020:RoAt}& \cofo & \cofotm & \cofolb & -- & \pmb{\checkmark}\\
    Catal{\'a}n \etal~\cite{S:2020:TaAa}& \cofo & \cofolb & -- & -- & \pmb{\checkmark}\\
    COEQT~\cite{zhang2016efficient}&\stfo & Tree & \dacache & \cofolb & \pmb{\checkmark}\\
    COO~\cite{saad1994sparskit}  & \stfo & Flat &-&-& \pmb{\checkmark}\\
    COOCOO~\cite{imecek:2012:SSMS:HPCC}& \stfo & Blocked &-&-& \pmb{\checkmark}\\
    CMSR~\cite{Kwiatkowska:2002:OSoL, Mehmoodthesis04} & \stfo & Flat & \mem & \stfolb & \pmb{\checkmark}\\
    CSB~\cite{Bulucc:2009:Psma}& \stfo & Blocked & \mem & -- & \pmb{\checkmark}\\
    CSR2~\cite{Bian:2020:CANF:CCGRID}& \cofo & \cofolb & Blocked & -- & \pmb{\checkmark}\\
    CSR~\cite{saad1994sparskit}  & \stfo & Flat&-&-& \pmb{\checkmark}\\
    CSRCOO~\cite{imecek:2012:SSMS:HPCC}&\stfo & \hierarchical&--& --& \pmb{\checkmark}\\
    CSRLen~\cite{Aktemur:2018:Asmm:CPE}& \cofo & \cofolb & \dacompiler & -- & \pmb{\checkmark}\\
    CSR-SIMD~\cite{Chen:2018:AeSc:CPE}& \stfo & \mem & \cofotm & -- & \pmb{\checkmark}\\
    CSX~\cite{Karakasis:2013:AECF:TPDS}& \stfo & \mem & \stfoma & -- & \pmb{\checkmark}\\
    DCSR~\cite{Willcock:2006:Asmc}& \stfo & \devdev & -- & -- & \pmb{\checkmark}\\
    DDD-Split~\cite{Yuan:2010:OSMV:HPCC}&\stfo & Hybrid &\bandw& --& \pmb{\checkmark}\\
    DDD-Naïve~\cite{Yuan:2010:OSMV:HPCC}&\stfo & Hybrid &\bandw& --& \pmb{\checkmark}\\
    DIA~\cite{saad1994sparskit} & \stfo & Flat & \stfolb& \stfoma & \pmb{\checkmark}\\
    Dvorsky and Kratky~\cite{dvorsky2004multi}& \stfo & Tree & -- & -- & --\\
    Elaforu \etal~\cite{DBLP:journals/corr/ElafrouGK15}& \auto & \mls & -- & -- & \pmb{\checkmark}\\
    ELLAPCK~\cite{GrimesKincaidYoung1979}  & \stfo & Flat&\stfolb& \stfoma& \pmb{\checkmark}\\
    HiSM~\cite{Stathis2004,Stathis:AHsm:IPDPS}& \stfo & \hierarchical&\mem& --& \pmb{\checkmark}\\
    GCRS/GCCS~\cite{Shaikh:2015:Essf:HPCSIM} & \stfo & Flat & -- & -- &\\
    JAD/JDS~\cite{Saad:1989:KSMo} & \stfo & Flat &\stfoma& -- & \pmb{\checkmark}\\
    MQT~\cite{Simecek:2012:MQFf:SYNASC}& \stfo & Tree & \mem & -- & --\\
    MSR~\cite{saad1994sparskit} & \stfo & Flat & \stfolb& \stfoma & \pmb{\checkmark}\\
    OSKI~\cite{Vuduc:2005:OAlo}& \cofo & \cofoat & \dacompiler  & -- & \pmb{\checkmark}\\
    RACE \etal~\cite{Alappat:2020:ARAC}& \cofo & \cofolb & \cofodr & -- & \pmb{\checkmark}\\
    RPCSR~\cite{Willcock:2006:Asmc}& \stfo & \devdev & -- & -- & \pmb{\checkmark}\\
    RSDIAG~\cite{Vuduc2003}  & \stfo & Sliced & \cofoat & -- & \pmb{\checkmark}\\
    SBSRS~\cite{ShahnazUsman2011}& \stfo & Blocked & \bandw & -- & \pmb{\checkmark}\\
    S-CSR~\cite{Guo:2010:OSDS,Guo:2013:Aots}& \stfo & Streams & \cofoma & -- & \pmb{\checkmark}\\
    SMAT~\cite{Li:2013:S}& \auto & \mls & \cofoat & -- & \pmb{\checkmark}\\
    SPARSITY~\cite{Im:2004:SOFf} & \cofo & \cofoat & -- & -- & \pmb{\checkmark}\\
    TJDS/TJAD~\cite{An:2003:AACS}& \stfo & Flat & \stfoma & \stfolb & \pmb{\checkmark}\\
    tpSpMV~\cite{Chen:2020:tAtl:INS} & \stfo & \dacache & \cofoma & \cofolb & \pmb{\checkmark}\\
    UBCSR~\cite{Vuduc:2005:Fsmm}& \stfo & \singlef & Blocked & -- & \pmb{\checkmark}\\
    Venkat \etal~\cite{An:2016:AWPf:SC}& \cofo & \cofoat & \dacompiler & -- & \pmb{\checkmark}\\
    ZAKI~\cite{Usman:2019:ZASM:S}& \auto & \mls & Blocked & \stfotm & \pmb{\checkmark}\\
    ZAKI+\cite{Usman:2019:ZAML:ACCESS}& \auto & \mls & Blocked & \stfotm & \pmb{\checkmark}\\
    Zhang \etal~\cite{Zhang:2018:VPSM}& \cofo & \cofotm & \cofolb & Sliced & \pmb{\checkmark}\\
  \end{longtable}
\end{footnotesize}
\normalsize

\subsection{Summary and Analysis (CPUs)}\label{sec:cpu.summary}
In this section, we have reviewed the notable SpMV works proposed for CPUs, summarised in Table~\ref{tabcpu.summ}. 
The structure of the table and the information about the taxonomies is similar to Table~\ref{tab:gpu:summ}, refer to Section~\ref{sec:gpu:summary} [Page~\pageref{sec:gpu:summary}] for explanation. 
The table shows that SpMV techniques for CPUs are rich in their taxonomy classes. 
There are thirty-two, eleven, and four works respectively focused on algorithmic, storage and auto-selective techniques. SpMV computations on CPUs generally are not bound for its FLOPS rather for storing large matrices in the CPU RAM. Therefore, storage has been a bigger challenge for CPU architectures compared to obtaining high performance from CPUs and this is visible in the table. Obtaining performance on CPUs is also very challenging and fascinating and hence we see a good number of interesting works. Auto-selective has a fewer techniques because it is a new trend and is expected to rise steeply in the near future due to the recent surge in the use of machine learning.

\section{SpMV Techniques on FPGA{s}}\label{sec:fpga}
Most of the works using FPGA provide an improvement to the FPGA architecture to improve SpMV. Majority of these works use the CSR or CSC formats for sparse matrix storage. Very few works have proposed new sparse matrix storage schemes. The proposed architectures can be classified into two broad categories, (1) each parallel PE (Processing Elements) computes a single element of vector $y$, and (2) each PE computes several elements of $y$ in a sequential manner. The first kind of architectures generally use reduction circuits.

\subsection{Storage-focused (Stfo) Techniques on FPGAs}\label{sec:fpga.stfo}

\subsubsection{Structural}\label{sec:fpga.stfo.struct}
\begin{description}[leftmargin=0pt, itemsep=0.5em, topsep=0.5pt]
\item[Blocked.]\label{sec:fpga.stfo.struct.bl}
Smailbegovic \etal~\cite{SmailbegovicGaydadjievVassiliadis2005} extended Block Row Compressed Format~\cite{Vassiliadis:2002:BBCS} and proposed the SPBCRSx format (extended Sparse Block Compressed Row Storage) to store sparse matrices on FPGAs. The scheme stores a matrix element and its index as a pair that, for a matrix row, are linked together to form a linked list.  
Other schemes include~\cite{Li:2016:Adld}, and 
~\cite{Wu:2013:HAft:TCSII}.
\end{description}

\subsubsection{Compression-Based}\label{sec:fpga.stfo.comp}
\begin{description}[leftmargin=0pt, itemsep=0.5em, topsep=0.5pt]
\item[Bandwidth-Specific.]\label{sec:fpga.stfo.comp.bw}
Kestur \etal~\cite{Kestur:2012:TaUF:FCCM} proposed a set of storage schemes named Bit vector formats, including Bit Vector (BV), Compressed Bit-Vector (CBV), and Compressed variable length Bit-Vector (CVBV). BV only stores zeroes and ones depending on the nonzero values of the matrix. CBV is a BV scheme with run-length encoding. In CVBV, the run-length encoding can be stored using variable-length non-zero rows, which offers storage savings. 

\item[Memory-Specific]\label{sec:fpga.stfo.comp.mem}
Boland and Constantinides~\cite{Bol:2011:Ombu} proposed circuit architectures for performing MVM that exploit symmetrical and banded matrix structures. The proposed MVM is aimed for on-chip buffering of data and hence is supposed to reduce the RAM usage on FPGAs for symmetric matrices. 
Grigoras \etal~\cite{Grigoras:2015:ASoF:FCCM} propose the BCSRVI scheme for compute SpMV on FPGAs. The scheme is based on CSRVI~\cite{Kourtis:2008:ItPo:ICPP} (proposed for CPUs) that does not store the redundant $\nnz$s. Due to the FPGA memory limitations, they restrict CSRVI such that only the k most frequently occurring values in the nonzero array are stored. 
\end{description}

\subsection{Computation-focused (Cofo) Performance Techniques for FPGAs}\label{sec:fpgacofo}

\subsubsection{Algorithmic}\label{sec:fpga.cofo.algo}

Zhou and Prasanna~\cite{Zhuo:2005:SMmo} claimed to have reported (in 2005) the very first implementation of SpMVs on FPGAs (see also deLorimier and DeHon~\cite{deLorimier:2005:Fsmm}). The proposed architecture is capable of performing multiple floating point operations as well as I/O operations in parallel and without any assumptions on the matrix sparsity structure. It consists of a tree of binary operators. In the tree each leaf node is considered as a floating point multiplier and the nodes before that are floating point adders. The output from the root node is fed to a reduction circuit which collects the intermediate results. For larger matrices a block based SpMV is also discussed. deLorimier and DeHon~\cite{deLorimier:2005:Fsmm} reported CSR-based SpMV on FPGAs by partitioning the dot products across multiple PEs. The entire computation is the set of dot products between the vector and the matrix rows (this work appeared in the same year and conference as that of Zhou and Prasanna~\cite{Zhuo:2005:SMmo}).
Bakos and Nagar~\cite{Bakos:2009:EMSt:FCCM}, for CG computations, exploit the symmetry of matrices in order to compute both the top triangle and bottom triangle of the input matrix in parallel.
The SpMV architecture is divided into two sections, the first performs computation for the upper triangle, including the diagonal, and the second for the lower triangle which are combined to produce the final results.
Kuzmanov and Taouil~\cite{Kuzmanov:2009:Rsmm:FPT} combined sparse MVM with Dense MVM on FPGAs. They analyze the boundary conditions when both the sparse and dense matrices provide the same performance, and based on this analysis, developed a system that improves the performance. The proposed design supports multiple operations, particularly: SpMV, dense MVM, and dense MMM.
Nechma \etal~\cite{Nechma:2010:Psms:ISCAS} used static pivoting and symbolic analysis to compute an accurate task-flow graph that improves the floating point performance of the system. The authors address direct linear solvers, particularly LU decomposition on sparse matrices that arise in SPICE circuit simulation. The CSR storage format is used to spread column entries over multiple PEs to facilitate loop unrolling.
Wu \etal~\cite{Wu:2013:HAft:TCSII} proposed a CG algorithm for solving Linear systems on FPGAS to reduce the zero-padding and remove the BRAM constraints for matrices of arbitrary sizes and sparsity patterns. The input matrix is divided into blocks and stored in the off-chip memory while the blocks of vectors $x$ and $y$ are stored in the internal memory. 
Pinhao \etal~\cite{Pinhao:2015:SMMo:DSD} proposed a method to load-balance SpMV on PEs in FPGAs with each PE being assigned a single row in a round robin fashion. They use CSC (Compressed Sparse Column) to store the sparse matrices to allow coalesced memory access for the vector. External memory is used to store the input data and the results are written back to the external memory.

Grigoras \etal~\cite{Grigoras:2016:OSMV:FPL} proposed an architecture and an automated customization method to detect and optimize the architecture for block diagonal sparse matrices. They proposed a performance model to tune the  proposed architecture for matrices and reduce the BRAM resource utilization on FPGAs by as much as 10 times.

Li et al.~\cite{Li:2016:Adld} proposed a data locality-aware design framework for FPGA-based SpMV acceleration. They also developed a distributed architecture composed of PEs to improve the computation parallelism. Moreover, they developed a locality-aware clustering technique for conventional sparse matrix compression formats. Consequently, a sparse matrix is reorganized into small matrix blocks, each of which has regularized memory accesses.
Sadi \etal~\cite{Sadi:2019:ESOf} proposed an algorithm co-optimized scalable hardware architecture aimed to efficiently perform SpMV on large and highly sparse matrices using smaller on-chip fast memory than existing solutions.  
Jain \etal~\cite{Jain:2020:ADAf:FPL} proposed ``sparstition'', a partitioning scheme for SpMV based on a High-level synthesis (HLS) based hardware pipelined design to allow the matrix size to be limited only by the size of the off-chip memory (DRAM) and not by the available on-chip memory (BRAMs). 
Parravicini \etal~\cite{Parravicini:2021:ArsS} proposed a streaming implementation of COO-based SpMV Personalized PageRank on FPGA that leverages data-flow computation and reduced precision fixed-point arithmetic.  
Other schemes include~\cite{Dorrance:2014:Assm},~\cite{Grigoras:2016:OSMV:FPL},~\cite{Li:2016:Adld}, and SPBCRSx~\cite{SmailbegovicGaydadjievVassiliadis2005}.

\begin{description}[leftmargin=0pt, itemsep=0.5em]
\item[Buffering.]\label{sec:fpga.cofo.buff}
The works in this category include~\cite{Wu:2013:HAft:TCSII},~\cite{Dorrance:2014:Assm},~\cite{Pinhao:2015:SMMo:DSD}, and~\cite{Grigoras:2016:OSMV:FPL}.

\item[Reduction.]\label{sec:fpga.cofo.reduce}
Prasanna and Morris~\cite{Prasanna:2007:SMCo:MC} proposed a CG solver and Jacobi solver for FPGAs. The proposed scheme performs large parallel reductions by breaking it into a sequence of smaller reductions. This is aimed to perform without stalling the pipeline or imposing unreasonable buffer requirements. 
Other works include~\cite{Zhuo:2005:SMmo}.

\item[Data-Reuse.]\label{sec:fpga.cofo.datareuse}
The works in this category include~\cite{Bakos:2009:EMSt:FCCM}.

\item[Memory-Alignment.]\label{sec:fpga.cofo.memalign}
Dorrance \etal~\cite{Dorrance:2014:Assm} proposed a new SpMV architecture in which each entire column of the vector $y$ is computed using column-wise vector addition of $A$  by $x$.
Other works include~\cite{Pinhao:2015:SMMo:DSD}.
\end{description}

\begin{footnotesize}
  \begin{table}[h!]
  \centering
    \footnotesize
    \caption{Classifications of SpMV Techniques (FPGAs)\label{tabfpga.summ}}
    \begin{tabular}[t!]{@{}llp{2.3cm}p{2.3cm}p{2.3cm}@{}}
      \toprule
      \multirow{2}{*}{Name} & \multicolumn{4}{c}{Taxonomy}\\
      \cmidrule(lr){2-5} & Class & Primary & Secondary&Tertiary\\
      \midrule 
      Bakos and Nagar~\cite{Bakos:2009:EMSt:FCCM}& \cofo & \cofolb & \cofodr & --\\
      Boland and Constantinides~\cite{Bol:2011:Ombu}& \stfo & \mem & \cofobuf & \devdev\\
      BV~\cite{Kestur:2012:TaUF:FCCM}   & \stfo & \bandw & -- & --\\
      CBV~\cite{Kestur:2012:TaUF:FCCM}  & \stfo & \bandw & -- & --\\
      CVBV~\cite{Kestur:2012:TaUF:FCCM} & \stfo & \bandw & -- & --\\
      deLorimier and DeHon~\cite{deLorimier:2005:Fsmm}& \cofo & \cofolb & -- & --\\
      Dorrance \etal~\cite{Dorrance:2014:Assm}& \cofo & \cofoma & \cofobuf & \mem\\
      Grigoras \etal~\cite{Grigoras:2015:ASoF:FCCM}& \stfo & \mem & -- & --\\
      Grigoras \etal~\cite{Grigoras:2016:OSMV:FPL}& \cofo & \cofoat & \cofolb & \cofobuf\\
      Li \etal~\cite{Li:2016:Adld}& \cofo & \cofodl & \cofolb & Blocked\\
      Kuzmanov and Taouil~\cite{Kuzmanov:2009:Rsmm:FPT}& \cofo & \cofolb & -- & --\\
      Nechma \etal~\cite{Nechma:2010:Psms:ISCAS}& \cofo & \cofolb & -- & --\\ 
      Pinhao \etal~\cite{Pinhao:2015:SMMo:DSD}& \cofo & \cofolb & \cofoma & \cofobuf\\
      Prasanna and Morris~\cite{Prasanna:2007:SMCo:MC}& \cofo & \cofor & \cofolb & --\\
      SPBCRSx~\cite{SmailbegovicGaydadjievVassiliadis2005} & \stfo & Blocked & \mem & --\\
      Wu \etal~\cite{Wu:2013:HAft:TCSII}& \cofo & \cofolb & Blocked & \cofobuf\\
    Zhou and Prasanna~\cite{Zhuo:2005:SMmo}& \cofo & \cofolb & \cofor & --\\
      \bottomrule
    \end{tabular}
  \end{table}
\end{footnotesize}
\normalsize

\subsection{Summary and Analysis (FPGAs)}\label{sec:fpga:summary}
In this section, we have reviewed the various SpMV works proposed for FPGAs. 
Table~\ref{tabfpga.summ} gives a summary of the discussed FPGA techniques. 
The structure of the table and the information about the taxonomies is similar to Table~\ref{tab:gpu:summ} (Section~\ref{sec:gpu:summary}, Page~\pageref{sec:gpu:summary}).
Recent works address the challenges faced by SpMV on FPGAs including low bandwidth utilization, limited on-chip memory, low compute occupancy, and timing closure on multi-die heterogeneous fabrics. The table shows that the majority of FPGA works are focussed on improving the algorithmic aspects (Cofo) as opposed to the storage aspects. It is understandable because FPGAs allow configurations of the compute hardware. Moreover, all the Stfo schemes are focussed on compression due to the limited RAM with FPGAs. 

\section{SpMV Techniques for MICs}\label{sec:mic}

\subsection{Storage-focused (Stfo) Techniques (MICs)}\label{sec:mic:stfo}

\subsubsection{Structural Techniques}\label{sec:mic:stfo:structural}

Liu \etal~\cite{Liu:2013:Esmm} proposed an efficient SpMV on Intel Xeon Phi (MIC) coprocessors by developing a new sparse storage format called ELLPACK Sparse Block (ESB). ESB is an extension of the ELLPACK format, developed to address the performance problems related to CSR-based SpMV on MICs.In addition to the regular ELLPACK format, column blocking is used to improve locality of memory accesses. Moreover, in each of the column, a finite window sort is applied to improve the SIMD efficiency. To reduce the memory-bandwidth requirements, the authors encode the indices of the nonzeroes into a bit array. However, the use of bit array degrades the performance for some matrices.
Saule \etal~\cite{Saule:2014:PEoS} reported performance results of CSR on Intel Xeon Phi. They used Reverse Cuthill-McKee (RCM) algorithm to reorder the matrix rows. 
Kreutzer \etal~\cite{Kreutzer:2014:AUSM} proposed SELL-C-$\sigma$ that could be used on both Intel MICs and GPUs. This scheme has been discussed in Section~\ref{sec:gpu:stfo:structural}.
Liu and Vinter~\cite{liu2015csr5} proposed CSR5 that could be used on GPUs, CPUs, and MICs. It has already been discussed in Section~\ref{sec:gpu:stfo:structural}.  
Tang \etal~\cite{Tang:2015:Oaas:CGO} proposed Vectorized Hybrid COO+CSR (VHCC) to improve the performance and data locality on MICs. 
An input matrix is partitioned using a 2D jagged blocking technique and tiled to improve the data locality. Vectorized sum computations are performed on these matrices using SIMD. The aim is to efficiently utilize the VPU (Vector Processing Unit) in MIC architectures.The nonzeros are combined together and are divided vertically into equal panels. Each panel is further divided into blocks such that the total number of nonzeros in each block are the same. Each coprocessor is assigned to one block. Each block is further divided into a vector of eight nonzeros. Each element in the vector is a tuple of data values and column indices. A vector pointer and a row pointer are also used to keep track of the vectors and the rows. The prefix sum segmented operation is further used to compute SpMV. The proposed scheme reports good performance for irregular matrices, but for regular matrices the performance gain is not significant.
Yzelman~\cite{Yzelman:2015:Gvfs} extended their earlier work~\cite{Yzelman:2014:HSfP:TPDS} (aimed for CPUs) and proposed a compressed vectorized BICRS format (VC-BICRS), which is based on BCSR, Sliced ELL, and segmented scan based techniques. The proposed technique performs sparse blocking and compression. They show that vector operations can improve performance on MIC based processors. The proposed technique performs well with structured matrices but falls behind on unstructured matrices as compared to GPUs.

\subsubsection{Compression-based Techniques (MICs)}\label{sec:mic:stfo:cb}

Yan \etal~\cite{Zhang:2016:ACSF} extended to run on both GPU and MIC architectures their earlier work on Blocked Compressed COO (BCCOO) and Block Compressed COO plus (BCCOO+) schemes that was presented in~\cite{YanLiZhangEtAl2014} and was only focused on GPUs. 

\begin{footnotesize}
\begin{table}
  \centering
    \caption{Summary and Classification of SpMV Techniques (MICs)}\label{tabmic.summ}
    \begin{tabular}[t!]{@{}llllll@{}}
      \toprule
\multirow{2}{*}{Name} & \multicolumn{4}{c}{Taxonomy} &  \multirow{2}{*}{SpMV}\\
      \cmidrule(lr){2-5} & Class & Primary & Secondary&Tertiary\\
      \midrule Alyahya \etal~\cite{Alyahya:2018:PSMV}  & \cofo & \cofolb& -- & -- & \pmb{\checkmark}\\
      Alyahya \etal ~\cite{Alyahya:2019:PISo} & \cofo & \cofolb & -- & -- & \pmb{\checkmark}\\
      Alzahrani \etal~\cite{Alzahrani:2018:PEoJ} & \cofo & \cofolb & -- & -- & \pmb{\checkmark}\\
      BCCOO~\cite{YanLiZhangEtAl2014} & \stfo &\mem& Multi-Format&Sliced& \pmb{\checkmark}\\
      BCCOO+~\cite{YanLiZhangEtAl2014} & \stfo &\mem& Multi-Format&Sliced& \pmb{\checkmark}\\
      CSR5~\cite{liu2015csr5} &\stfo  & Blocked&\stfolb&Single-Format& \pmb{\checkmark}\\
      ESB~\cite{Liu:2013:Esmm} & \stfo & Blocked & \stfocm & Bandwidth & \pmb{\checkmark}\\
      Saule \etal~\cite{Saule:2014:PEoS}& \stfo & Sorted & \stfolb & -- & \pmb{\checkmark}\\
      SELL-C-$\sigma$~\cite{Kreutzer:2014:AUSM} &\stfo & Sorted & \stfolb & \cofolb& \pmb{\checkmark}\\
      VC-BICRS~\cite{Yzelman:2015:Gvfs} & \stfo &{Blocked} & \bandw Bandwidth & -- &\pmb{\checkmark}\\
      VHCC~\cite{Tang:2015:Oaas:CGO} & \stfo & Blocked & \stfocm & -- & \pmb{\checkmark}\\
      \bottomrule
    \end{tabular}
  \end{table}
\end{footnotesize}
\normalsize

\subsection{Computation-focused (Cofo) Techniques (MICs)}\label{sec:mic:cofo}
Alyahya \etal~\cite{Alyahya:2018:PSMV} 
reported a parallel implementation of CSR-based SpMV on Intel MIC (Knights Corner (KNC)) using the offload programming model with OpenMP and compared its performance with the sequential version in terms of memory usage, execution and offloading times, and speedup. 
They extended their work in~\cite{Alyahya:2019:PISo} to implement Jacobi interative method for solving sparse linear equation systems using CSR and Modified Sparse Row (MSR) and comparing results of MIC with a 24-core CPU node.
A similar work is presented by Alzahrani \etal~\cite{Alzahrani:2018:PEoJ} with a different matrix suit and load balancing approach. 

\subsection{Summary and Analysis (MICs)}\label{sec:mic:summary}
In this section, we have reviewed the various SpMV works proposed for MICs. 
Table~\ref{tabmic.summ} gives a summary of the discussed MIC techniques. 
The structure of the table and the information about the taxonomies is similar to Table~\ref{tab:gpu:summ} (Section~\ref{sec:gpu:summary}, Page~\pageref{sec:gpu:summary}). 
The table shows that the majority of MIC works are focussed on improving the storage aspects. The majority of the works focus on utilizing the 512 bits vector SIMD lane of MICs with coalesced memory access. Unlike GPUs MICs does not have a texture cache which worsens the irregular memory access and causes a load imbalance. Register blocking can in addition lead to lower SIMD efficiency. Hence, there is a potential of exploring storage formats.

\section{SpMV Techniques on Heterogeneous Architectures}\label{sec:hetro}

Researchers have attempted to improve the SpMV performance on heterogeneous architectures and these are reviewed in this section without aiming to be exhaustive. We include the works here which either partition the SpMV computations over multiple architecture or they could be executed over multiple architectures without modifications.
Indarapu \etal~\cite{Indarapu:2014:Aawa} proposed three SpMV algorithms that simultaneously utilize both CPU and GPU. They try to determine effective mechanisms to allocate an appropriate amount of work to CPU and GPU. Cardellini et al.~\cite{Barbieri:2012:DPSC} applied object-oriented programming (OOP) design patterns to develop a user interface and high-level scientific software involving sparse matrices that allows efficient use of CPU/GPU platforms using the PSBLAS library. Yang \etal~\cite{Yang:2015:POUP:TC} developed a sparse matrix partitioning method to improve SpMV performance on CPU-GPU heterogeneous systems. The method obtains dense blocks by analyzing the probability distribution of nonzeros in a sparse matrix to be executed on CPU and GPU.  
The CSR5 scheme~\cite{liu2015csr5} that has been discussed earlier also falls in the heterogeneous category because provides consistent performance on GPUs, CPUs, and MICs for the selected dataset.  
Yang \etal~\cite{Yang:2017:Ahcm:JPDC} proposed a scheme based on HYB that executes the ELL part of the matrix on GPU and COO part on CPU, concurrently. The ELL part is replaced with DIA based on the matrix structure. Benatia \etal~\cite{Benatia:2019:Smpf} use Support Vector Regression (SVR) to optimize the horizontal partitioning of sparse matrices into row-blocks (slices), select the best format (CSR, ELL, or HYB) to store each slice, and map the partitions onto GPU and CPU for execution.
Yan \etal~\cite{Zhang:2016:ACSF} extended their earlier work on BCCOO and BCCOO+ schemes~\cite{YanLiZhangEtAl2014}, which were focussed on GPUs alone to run on both GPU and MIC architectures. 
Alahmadi \etal~\cite{AlAhmadi:2020:PAoS:ELECTRONICS} propose a technique to improve performance of HYB by executing all the reduce operations (i.e.\ atomic add) on CPU\@. Originally, all HYB computations are entirely executed on GPU where atomic operations could slow down the GPU performance due to the need for thread synchronisations.

Various software libraries aim at designing linear algebra algorithms and frameworks for hybrid multicore and multi-GPU systems. They include MAGMA (multicore + multi-GPU systems)~\cite{Yamazaki:2014:IPDPS}, PETSc (MPI + GPU)~\cite{Zhang:2022:TPDS}, Trilinos module Tpetra (MPI + GPU)~\cite{trilinos:website}, Kokkos (MPI + GPU)~\cite{Trott:2022:TPDS}, and GHOST~\cite{Kreutzer:2016:S} (MPI + GPU).

\begin{footnotesize}
\begin{table}
  \centering
    \caption{Summary and Classification of SpMV Performance Techniques (Heterogeneous Architectures)}\label{tabhetero.summ}
      \begin{tabular}[t!]{@{}llllll@{}}
      \toprule
\multirow{2}{*}{Name} & \multicolumn{4}{c}{Taxonomy} &  \multirow{2}{*}{Platform}\\
      \cmidrule(lr){2-5} & Class & Primary & Secondary&Tertiary\\
      \midrule Indarapu \etal~\cite{Indarapu:2014:Aawa} & \cofo &\cofolb & -- & -- & CPU+GPU\\
      Cardellini et al.~\cite{Barbieri:2012:DPSC} & \cofo & -- & -- & -- & CPU+GPU\\
      Yang \etal~\cite{Yang:2015:POUP:TC}       & \cofo & \cofolb & Blocked & -- &CPU+GPU\\
      CSR5~\cite{liu2015csr5} &\stfo  & Blocked&\stfolb&Single-Format& CPU/GPU/MIC\\
      Yang \etal~\cite{Yang:2017:Ahcm:JPDC} & \cofo & Multi-Format & \cofolb & -- & CPU+GPU\\
      Benatia \etal~\cite{Benatia:2019:Smpf}& \cofo & \mls & Multi-Format & \cofolb & CPU+GPU\\
      Yan \etal~\cite{Zhang:2016:ACSF} & \cofo & Memory-specific & Multi-Format & Sliced & CPU/GPU/MIC\\
      Alahmadi \etal~\cite{AlAhmadi:2020:PAoS:ELECTRONICS} & \cofo & Multi-Format & \cofolb & -- & CPU+GPU\\
      \bottomrule
    \end{tabular}
  \end{table}
\end{footnotesize}
\normalsize

\subsection{Summary and Analysis (Heterogeneous Architectures)}\label{sec:hetro:summary}
Table~\ref{tabhetero.summ} summarises the classification of the schemes aimed at heterogeneous architectures in the same manner as we have presented earlier for individual architectures (see Section~\ref{sec:gpu:summary}, Page~\pageref{sec:gpu:summary}). The only difference is the last column, which gives the specific device architectures that the scheme is able to execute on. The symbol "+" between the device names indicates that the scheme parallelizes the computations on the multiple devices. The symbol "/" implies that the scheme can run on multiple architectures separately and is not parallelized to run concurrently. 
Note in Table~\ref{tabhetero.summ} that all except one schemes are focused on algorithmic aspect (\cofo). This is understandable because heterogeneous computing is typically focussed on distributing computation tasks. The number of SpMV schemes for heterogeneous architectures are expected to increase in the future due to the increase in the heterogeneity of computing architectures.

\begin{figure}
  \centerline{\includegraphics[scale=0.65]{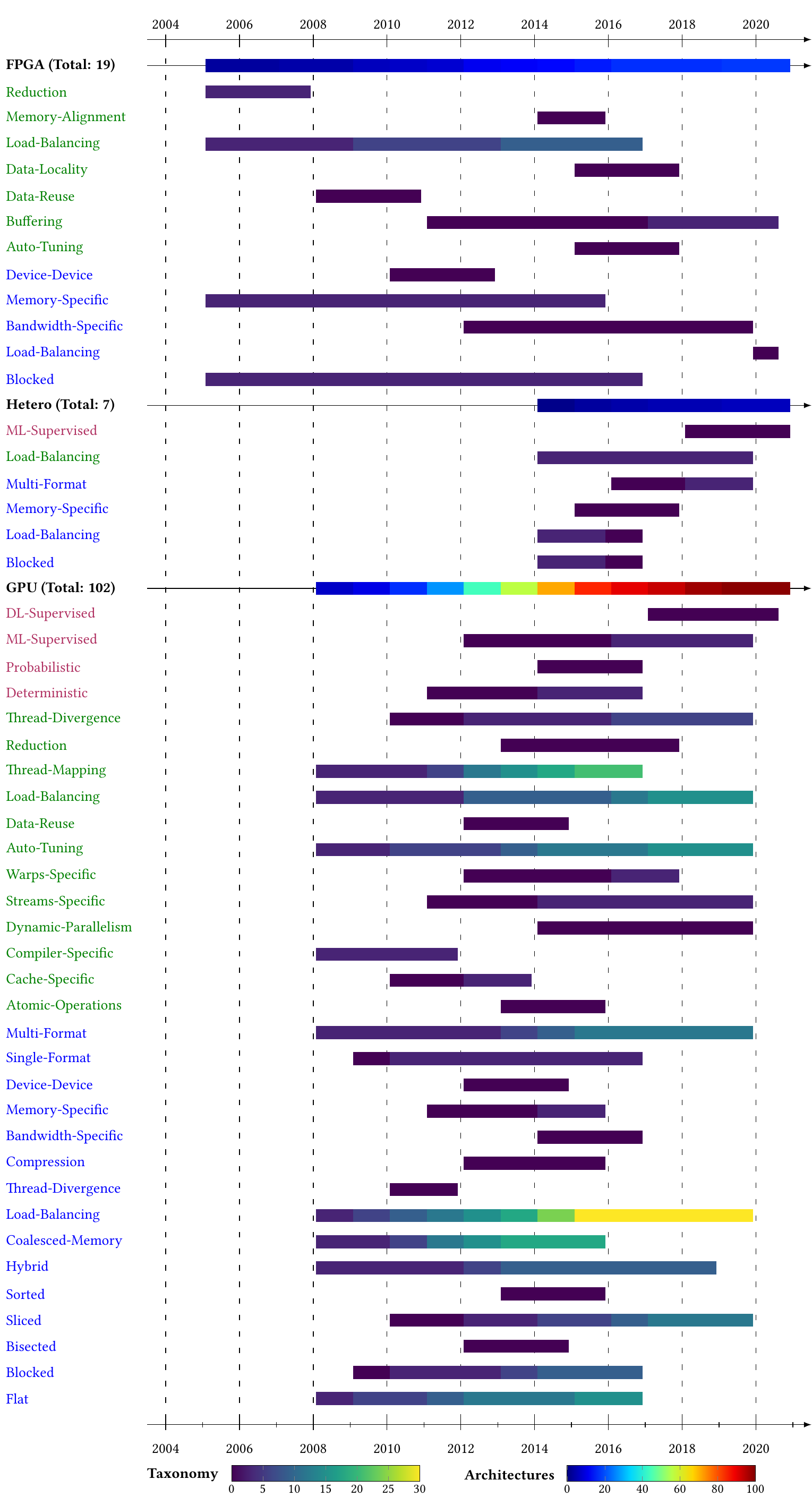}}
  \caption{SpMV schemes: taxonomies by architectures}\label{fig:taxo-gpu-hetro}
\end{figure}
\begin{figure}
  \centerline{\includegraphics[scale=0.8]{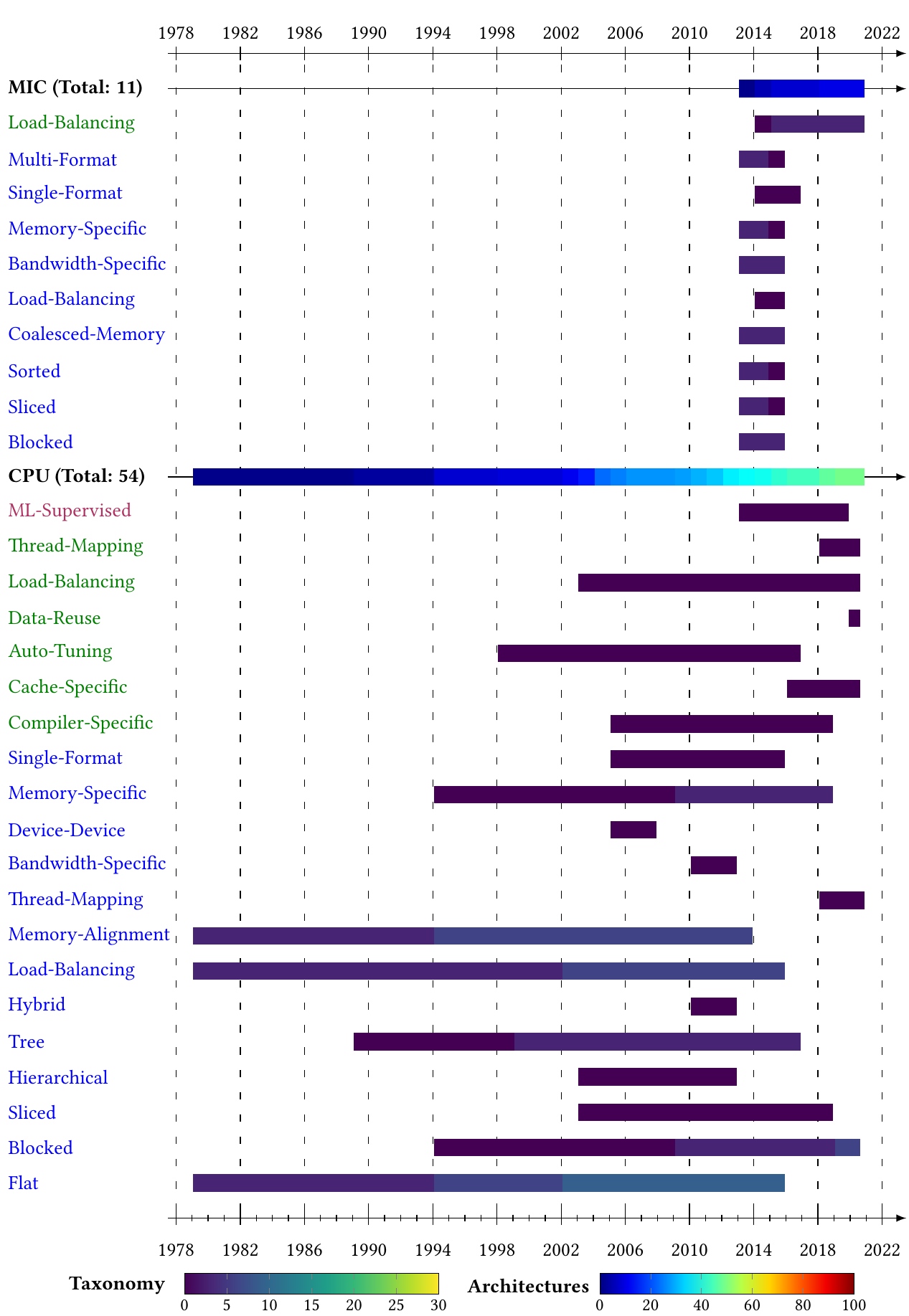}}
  \caption{SpMV schemes: taxonomies by architectures}\label{fig:taxo-cpu-mic}
\end{figure}

\section{Conclusions and Future Research Directions}\label{sec:conclusion}

In this paper, we reviewed the performance enhancement techniques for SpMV computations on major high performance computing processor architectures (CPU, GPU, FPGA, MIC, and heterogeneous architectures). The review also includes works on the iterative solution of large sparse linear equation systems that have a focus on SpMV performance. Based on the review we developed a multi-architectural taxonomy of SpMV research and classified the literature accordingly. While we reviewed the works on all major processor architectures, our main focus was on GPU architectures. To the best of our knowledge, no such taxonomy or detailed review of SpMV works on multiple architectures exists. 

The literature review presented in this paper shows the sheer richness of the proposed techniques for SpMV computation, and it is expected due to the significance of the research in SpMV and its being among the seven dwarfs.  
A great deal of research has gone into devising matrix storage and SpMV compute schemes due to the complexity of both the problem space (extreme variations in the sparsity structure and size of matrices) and the target processor architectures (multi- and many-cores with complex compute and memory settings). 
The Stfo techniques are very rich in their approaches including a range of structural, speed-enhancement, compression-based, and configurable methods to improve SpMV performance. 
The Cofo techniques have focussed on improving SpMV performance by exploiting device architectural features and addressing various performance features such as auto-tuning, thread-mapping, etc.
The Sefo techniques are based on analytical or AI-based approaches to select the optimal storage scheme or a compute kernel for a given matrix. Sefo schemes based on PP models are hard to build and extend. They make many assumptions and hence do not provide high performance. ML-based Sefo techniques are easy to develop and are able to provide high performance.   
We reviewed around 102 GPU-based techniques (storage technique as well as architectural optimizations), 30 FPGA and MIC-based performance improvement techniques for SpMV, 7 heterogeneous techniques, and 54 CPU-based implementations (Serial and parallel). A total of 190 schemes and techniques have been discussed and analyzed in the article. We do not claim to have done an exhaustive review of the schemes on all processor architectures so these numbers cannot be taken as a measure for SpMV works on CPU, MIC, and FPGA architectures.

Figure~\ref{fig:taxo-gpu-hetro} illustrates the development of discussed SpMV techniques on GPU, heterogeneous architectures, and FPGAs over time. Similarly, Figure~\ref{fig:taxo-cpu-mic} shows the development of SpMV techniques on MICs and CPUs over time. The GPU techniques have been covered in detail, whereas the other architectures have been covered to develop a common taxonomy. For GPUs, the works focusing on Stfo (60) are more in numbers than that of Cofo (28). There is a recent surge in ``Auto-Selective'' SpMV techniques of which there are 14 works in the past four years. All techniques other than ``Auto-Selective'' techniques do not cater to all matrix structures. However, ``Auto-Selective'' techniques yet have to make progress in improving the accuracy of storage detection. There still exist many open issues and problems and thee are discussed below in relation to future research directions. 

\subsection{Future Research Directions}\label{sec:future}
This century has seen unprecedented advances in computational devices and infrastructures. Multi-core CPUs, GPUs, FPGAs, and MICs, all have shown tremendous improvements in providing remarkable computational capacities for studying complex intractable problems. Solid state secondary memories and lower-delay, higher-bandwidth, relatively inexpensive memory hierarchies have empowered solving extraordinarily large problems. Lightening fast inter-node and host-to-device networks have paved the way for the exascale computing era. However, a major challenge is to bridge the hardware-software gap, utilize the available resources to the optimum, and translate the massive raw computational power into intelligence and information~\cite{Giles:2014:Tihc:RSTA}. These and many other transformational developments have given rise to new frontiers in SpMV computations. In this context, some potential directions are given below.  

\begin{itemize}
  \item AI including machine learning and deep learning has opened up new frontiers for virtually everything in our lives. It could also have a radical impact on SpMV design. For example, AI could be used to predict the best SpMV storage scheme and/or computation kernel for a given sparse matrix (\ie~Selective techniques; see Section~\ref{sec:gpu:select}). Dynamic runtime storage and computational optimizations could be developed and deployed. A challenge in this context would be to develop effective feature extraction and featureless AI schemes. AI-based approaches for matrix operations have unimaginable potential and could be evidenced in the recent work, ``discovering faster matrix multiplication algorithms with reinforcement learning"~\cite{10.1038/s41586-022-05172-4}.
  \item Substantial work is needed on bridging the hardware-software gap and utilizing the immense raw computational power available today in multicore (CPU), manycore (GPU, MIC), and programmable devices (FPGAs). Given the complexity of emerging devices, high-performance compilers and auto-parallelisation software are needed to reduce efforts on the human part. AI could help make significant advancements in this area. 
  \item Despite the relatively short history of GPUs (10 years or so compared to over 50 years of history of CPUs), the number and rich variety of proposed SpMV techniques on GPUs compared to CPUs, MIC, and FPGAs, show great potential for future developments. We believe this is due to two reasons. Linear algebra suits well to GPU architectures, naturally leading researchers to propose new schemes on GPUs for sparse and dense algebra. Secondly, GPU architectures are pretty configurable and have been evolving with new features which are being used by researchers to improve and propose new schemes. FPGAs are very configurable but require much more effort than GPUs. The various applications for FPGAs have been very domain specific. MICs have seen limited works on SpMVs to date, which we believe is due to their limited success in memory bandwidth-bound applications such as SpMV. MIC architectures are evolving and we expect new SpMV developments to emerge in the future. 
  \item SpMV research on heterogeneous architectures is in its infancy. Further work is needed in this direction such as multi-GPUs, and CPUs with GPUs, MICs, and FPGAs. Faster communication technologies such as NVLink would accelerate these trends. 
  \item The emerging efforts on big data and HPC convergence would lead to data-driven high performance SpMV techniques.
  \item Energy-efficient computing has become a priority in the exascale era and this would be another dimension or direction in future SpMV designs. 
  \item There is a need for standardized performance metrics to evaluate and compare SpMV performance (see \eg{} Langr and Tvrdik~\cite{Langr:2016:ECfS:TPDS}). The community needs to put substantial efforts in this direction. See~\cite{Mohammed:2020:DAnd:S, AlAhmadi:2019:PCfS}, for instance, where a variety of methods and metrics for performance analysis, visualization, and comparison of SpMV tools are used.
  \item The convergence of big data, AI, and HPC~\cite{Reed:2015:Ecab, Asch:2018:Bdae, Usman:2019:ZAML:ACCESS, Usman:2019:BDaH}, and their integration with the cloud, fog, and edge computing is opening up many new opportunities~\cite{Mohammed:2020:URLt:APP, Janbi:2020:DAID:S, Janbi:2022:IARA:S}.
  \item Research in data locality for tightly- and loosely-coupled architectures would become increasingly important due to the rise in edge-enabled distributed applications~\cite{10.20944/preprints202211.0161.v1}. 
\end{itemize}

\section*{Acknowledgement}
The authors acknowledge with thanks the technical and financial support from the Deanship of Scientific Research (DSR) at the King Abdulaziz University (KAU), Jeddah, Saudi Arabia,
under Grant No. RG-10-611-38. The work carried out in this paper is supported by the HPC Center at the King
Abdulaziz University.

\bibliographystyle{unsrtnat}
\bibliography{abbrev.bib,spmv-survey.bib}

\end{document}